\documentclass[referee]{raa}            

\usepackage{graphicx,times}   
\usepackage[a4paper, total={6.5in, 9in}]{geometry}

\usepackage{natbib}
\usepackage{booktabs}
\usepackage{amssymb,amsmath}
\usepackage{adjustbox}
\usepackage{threeparttable}

\usepackage{longtable}
\usepackage[colorlinks=true, linkcolor=blue, citecolor=blue, urlcolor=blue]{hyperref}

\usepackage{threeparttablex}

\begin{document}

  \title{The LAMOST Spectroscopic Survey of Supergiants in M31 and M33}

   \volnopage{Vol.0 (20xx) No.0, 000--000}     
   \setcounter{page}{1}          
   \author{Hao Wu 
      \inst{1,2}
   \and Yang Huang
      \inst{3,4,8}
   \and Huawei Zhang
      \inst{1,2,8}
   \and Haibo Yuan
      \inst{5,6}
   \and Zhiying Huo
      \inst{4}
   \and Cheng Liu
      \inst{7}
      }

   \institute{Department of Astronomy, School of Physics, Peking University, Beijing 100871, China; zhanghw@pku.edu.cn\\
        \and
        Kavli Institute for Astronomy and Astrophysics, Peking University, Beijing 100871, China\\
        \and
        School of Astronomy and Space Science, University of Chinese Academy of Science, Beijing 100049, China; huangyang@ucas.ac.cn\\
        \and
        National Astronomical Observatories, Chinese Academy of Sciences, Beijing 100101, China\\
        \and
        Institute for Frontiers in Astronomy and Astrophysics, Beijing Normal University, Beijing 102206, China\\     
        \and
        Department of Astronomy, Beijing Normal University, Beijing 100871, China\\
        \and
        Department of Scientific Research, Beijing Planetarium, Beijing, 100044, China\\
        \and
        Corresponding Authors\\
\vs\no
   {\small Submitted in 2024 June; Accepted in 2024 November}}

\abstract{ We present systematic identifications of supergiants of M31/M33 based on LAMOST spectroscopic survey. Radial velocities of nearly 5000 photometrically selected M31/M33 supergiant candidates have been properly derived from the qualified spectra released in LAMOST DR10. By comparing their radial velocities with those predicted from the rotation curves of M31, as well as utilizing {\it Gaia} astrometric measurements to exclude foreground contaminations, 199 supergiant members in M31, including 168 `Rank1' and 31 `Rank2', have been successfully identified. This sample contains 62 blue supergiants (BSGs, all `Rank1'), 134 yellow supergiants (YSGs, 103 `Rank1' and 31 `Rank2') and 3 red supergiants (RSGs, all `Rank1'). For M33, we identify 84 supergiant members (56 `Rank1' and 28 `Rank2'), which includes 28 BSGs (all `Rank1'), 53 YSGs (25 `Rank1' and 28 `Rank2') and 3 RSGs (all `Rank1'). So far, this is one of the largest supergiant sample of M31/M33 with full optical wavelength coverage (3700 \textless $\lambda$ \textless 9100 \AA). This sample is valuable for understanding the star formation and stellar evolution under different environments. 
}

   \authorrunning{H. Wu, Y. Huang \& H.W. Zhang et al. }            
   \titlerunning{The LAMOST Spectroscopic Survey of Supergiants in M31 and M33 }  

   \maketitle

\section{Introduction}           
\label{sect:intro}
\begin{figure}
   \centering
   \includegraphics[scale=1]{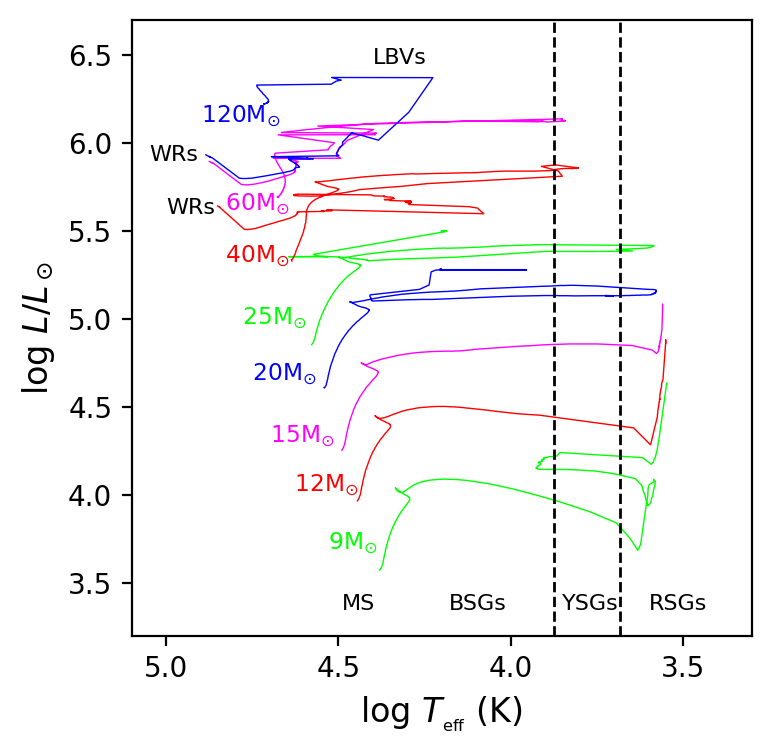}
   \caption{ The evolution tracks of massive stars ($M$ \textgreater 8 $M_{\odot}$) for solar metallicity (z = 0.014) assuming an initial rotation velocity of 40\% of the critical break-up speed taken from Geneva evolutionary tracks \citep{Ekstrom2012}. The evolution tracks with different initial masses are labeled with different colors. The two black vertical dashed lines represent the YSGs region, located in 4800 \textless $T_{\rm eff}$ \textless 7500 K. Other evolution phases (BSGs, RSGs, WRs and LBVs) are also labeled at their corresponding locations. }
   \label{fig:sg_hrd_eks}
   \end{figure}

Supergiants, as the evolution phases of massive stars (\textgreater 8 $M_{\odot}$), are extremely rare but of great interests because their importances in constraining the stellar evolutionary theory (e.g., \citealt{Massey2003,Massey2010,Massey2013,Massey2016}) and their profound feedback effects to the overall evolution of their host galaxies (e.g., \citealt{Oey2007}). Several different evolution stages after the termination of main sequence of massive stars can be classified according to their locations on the Hertzsprung-Russell (H-R) diagram (see Fig~\ref{fig:sg_hrd_eks}, where the evolution tracks are taken from \citealt{Ekstrom2012}). As shown in Fig~\ref{fig:sg_hrd_eks}, in the luminous blue range, blue supergiants (BSGs) are labeled, which are actually a mix of hot main sequence and other more evolved blue supergiants, including the evolved descendants of the most massive O-type stars: Wolf-Rayet stars (WRs). In addition, for the most luminous part of the blue region, the classical luminous blue variables (LBVs) phase will be the next stage of those most massive (e.g., $M$ \textgreater 30 -- 60 $M_{\odot}$) main sequence stars, which exhibit spectacular eruptions with several magnitudes enhanced visually and large mounts of material ejected (see, e.g., \citealt{Bohannan1997,Conti1997}). In the central part of the H-R diagram, the region within the two black dashed lines as marked in Fig~\ref{fig:sg_hrd_eks}, yellow supergiants (YSGs) are labeled, which are F/G-type supergiants with effective temperature ($T_{\rm eff}$) roughly from 4800 to 7500 K and luminosity log $L$/$L_{\odot}$ \textgreater 3.5. The YSG is a short-lived stage, which can either evolve from the main sequence massive stars or evolve back from the red supergiants (RSGs) for stars with initial mass between 8 and 30 $M_{\odot}$. Finally, the RSGs phases (K/M-type supergiants) locate in the cool (or red) region of the H-R diagram again as shown in Fig~\ref{fig:sg_hrd_eks}, which are relatively longer lived evolved descendants of moderately massive (8 -- 25 $M_{\odot}$) stars.
It is difficult to model the evolution tracks of massive stars, mainly due to the complicate mass loss process, which is still somewhat uncertain and lots of works (both theory and observation) are required to fully understand this process (e.g., \citealt{Massey2013,Smith2014}). The systematic search and study of supergiants in the Local Group (LG) of galaxies (perfect laboratories with a simple variables: chemical compositions) will provide important clues to constraining the theoretical models of stellar evolution. In addition, the studies of supergiants in the LG galaxies will also help us to understand their profound feedback effects that affect the evolution of these LG galaxies themselves (reviewed by \citealt{Oey2007}).

To systematic identify the supergiants in LG galaxies, \citet{Massey2006,Massey2007a} carried the Local Group Galaxies Surveys (LGGS) to obtain UBVRI photometry with precision better than 1 -- 2 percent for two spiral galaxies M31 and M33 and for seven irregular galaxies IC 10, NGC 6822, WLM, Sextans A, Sextans B, Pegasus and Phoenix.  With the precise photometry, one can select different types of supergiant candidates based on the color-magnitude or color-color diagrams. However, the foreground contamination hampers the selection of supergiants effectively through photometry only, especially for YSGs. In addition, it is difficult to obtain accurate physical properties for those supergiants with photometry information available alone. Fortunately, spectroscopy can solve the above problems neatly. Using the Hectospec spectrography equipped with 300 optical fibers mounted on MMT 6.5 m telescope, systematic identifications of YSGs in M31 and of YSGs/RSGs in M33 are carried by \citet{Drout2009,Drout2012}, respectively. Using the Hydra multi-object fiber spectrograph on WIYN 3.5 m telescope, RSGs are systematically searched in M31 by \citet{Massey2009}. Most recently, a more comprehensive search for supergiants in M31 and M33 are carried by \cite{Massey2016} by using the Hectospec spectrography mounted on MMT. By combing former efforts, a large catalog of 700/1200 supergiant candidates of M31/M33 is constructed by \cite{Massey2016}. 

Although significant progresses have been made in searching supergiants of M31/M33 in the past several years by Massey's group described above, the current supergiant sample of M31/M33 is still not complete because: 1) The number of photometric selected supergiant candidates is quite large and hard fully spectroscopically; 2) The wavelength coverage of current spectra is not covered with full optical range and thus not enough for further studies, such as metallicity estimates \citep{Liu2022}. 

The Large Sky Area Multi-Object Fiber Spectroscopic Telescope (LAMOST, also named the Guoshoujing Telescope; \citealt{Cui2012}) is a 4 meter quasi-meridian reflecting Schmidt telescope equipped with 4000 fibers, each of an angular diameter of 3\arcsec.3 projected on the sky, distributed in a circular field of view (FoV) of 5° in diameter. The LAMOST spectroscopic survey (\citealt{Zhao2012,Deng2012,Liu2014}) has huge advantage in systematic searching supergiants of M31/M33 given the fact that: 1) LAMOST is one of the telescope with highest rate of spectral acquisition. By the end of 2022, LAMOST has accumulated more than 21 million spectra, consisting of 11 million low resolution spectra and 10 million medium resolution spectra \citep{Yan2022}; 2) The FoV of LAMOST is near to the size of M31/M33; 3) The spectrographs used by LAMOST spectroscopic survey yield spectra covering the whole optical range (3700 \textless $\lambda$ \textless 9100 \AA)  with a moderate low resolution ({\it R} $\sim$ 1800). 

In this study, we present our systematic identification of supergiants in M31/M33 utilizing LAMOST DR10 data. Our approach involves the initial selection of photometric candidates through LGGS and {\it Gaia} photometry, which is described in Section~\ref{sect:select_obs}. Subsequently, the spectroscopy from LAMOST and astrometry from {\it Gaia} are used to select true members in Section~\ref{sect:identify}. In Section~\ref{sect:result}, we will show the results and make some discussions. Finally, a summary is presented in Section~\ref{sect:summary}.

\section{Target Selection and Observations}
\label{sect:select_obs}
As an extension of LAMOST Spectroscopic Survey of the Galactic Anticentre (LSS-GAC; \citealt{Liu2014}), the M31 and M33 areas (0$^\circ$ \textless RA \textless 30$^\circ$ and 25$^\circ$ \textless Dec \textless 50$^\circ$) were systematically observed by LAMOST. The detail survey strategy and target selections are descibed in \cite{Yuan2015}. Specially, potential luminous M31/M33 sources (such as the planetary nebulae, H II regions, globular clusters, supergiant stars) and background quasars were observed with high priorities. In this work, we utilized LAMOST DR10 LSR data, which is a collection of spectra obtained from October 2011 to June 2022. Here, our focus is on the supergiants of M31/M33 that mainly located in the disk region of the two galaxies. Therefore, hereafter the M31/M33 area refers to about 3/1.5 square degree around their centers (RA = 10$^\circ$.68458, Dec = 41$^\circ$.26875 for M31 and RA = 23$^\circ$.46208, Dec = 30$^\circ$.65994 for M33; \citealt{Jarrett2003}).

Here we introduce the detailed target selection of our supergiant candidates of M31/M33 in LAMOST DR10. First, the LGGS photometry catalogue is cross-matched with the LAMOST DR10, resulting in 1514 common sources for M31 and 981 ones for M33. Subsequently, by applying criteria from previous studies (mainly from Massey's group; e.g., \citealt{Massey2006,Massey2009,Drout2009,Drout2012}), the different types of supergiant candidates are selected as follows:
\begin{enumerate}
    \item
    BSG (including WR and LBV) candidates:
    \begin{equation}\label{eq:bsg}
        Q = (U-B) - 0.72 (B-V) \leq - 0.6.
    \end{equation}
   \item
    YSG candidates:
    \begin{equation}\label{eq:ysg1}
        U-B > - 0.4.
    \end{equation}
    \begin{equation}\label{eq:ysg2}
        0.0 \leq B-V \leq 1.5.
    \end{equation}
   \item
    RSG candidates:
    \begin{equation}\label{eq:rsg1}
        V-R \geq 0.85.
    \end{equation}
    \begin{equation}\label{eq:rsg2}
        B-V > - 1.6 (V-R)^2 + 4.18 (V-R) - 0.83.
    \end{equation}
\end{enumerate}
The greatest difficulty in selecting blue massive stars using photometry is the dust reddening. Though the negligible foreground extinction for entire M31 [{\it E(B $-$ V)} $=$ 0.06] and M33 [{\it E(B $-$ V)} $=$ 0.07] \citep{Van2000}, the inner dust of the two gas rich galaxies themselves will make those blue massive stars heavily and inhomogeneously reddened. The Johnson {\it Q}-index, as defined in Eq.~\ref{eq:bsg}, is a reddening-free indicator (at least for {\it Q} \textless $-$ 0.6) of intrinsic color. This index can work effectively in selecting blue massive supergiants \citep{Massey2016}. For YSGs, the photometric cuts in Eqs.~\ref{eq:ysg1}-\ref{eq:ysg2} only show their possible locations on color-color diagram and a large number of foreground contaminations also remain unavoidably. The identifications of the real YSGs will highly relied on the results from spectroscopy or astrometry. The foreground contamination is also a challenge of identifying RSGs in M31/M33.  But the color-color diagram (see Eq.~\ref{eq:rsg2}; \citealt{Massey1998,Drout2012}), {\it B $-$ V} and {\it V $-$ R}, can effectively distinguish RSGs and dwarfs, especially for {\it V $-$ R} \textgreater 0.6. This is simply because the {\it V $-$ R} is sensitive to effective temperature and {\it B $-$ V} is sensitive to both effective temperature and surface gravity.

The LGGS photometry has some caveats in selecting supergiant candidates: 1) The bright end of LGGS is not complete; 2) LGGS does not cover the entire M31/M33 sky region; 3) One or more bands of $UBVRI$ are missing for some sources in the LGGS catalog. To address the first issue, we incorporated objects from \cite{Magnier1992} at the bright end.  For the second and third cases, we utilized {\it Gaia} DR3 broadband $G/G_{\rm BP}/G_{\rm RP}$ photometry to select supergiant candidates using the criteria developed by \cite{Salomon2021}. Notably, these criteria merely represent the positions of different types of supergiants on the color-color diagram, and candidates selected through this way may suffer a substantial number of foreground contaminants.

In addition to above photometric criteria, three cuts related to LAMOST observations are applied. The first one requires {\it V} or {\it G} (for those without {\it V} photometry) magnitude to be brighter than 20 to match the limiting magnitude of LAMOST. The second one requires no nearby stars within 4\arcsec.00 (similar to the diameter of the LAMOST fiber) of the supergiant candidates, or if any, they should be at least 2 mag fainter, so as to exclude contaminants from close bright neighbors. The last cut requires the spectral signal-to-noise ratio (SNR) to be greater than 5, ensuring the quality of the spectrum for subsequent analysis. Finally, we excluded the contaminants from M31/M33's extended sources, i.e., globular clusters, planetary nebulae, and H II regions, by cross-matching with catalogs from \cite{Chen2015}, the M31 Revised Bologna Clusters and Candidates Catalog (\citealt{RBC2004,RBCV52014}) and \cite{Sarajedini2007}, as well as catalogs from \cite{Yuan2010}, \cite{Azimlu2011}, \cite{Sanders2012}, \cite{Zhang2020}, \cite{Hodge1999} and \cite{Ciardullo2004}.

The number of M31/M33 supergiant candidates is summarized in Tables~\ref{tab:targets} and ~\ref{tab:targets_gaia}. The number of RSG candidates is very small due to their intrinsic faint luminosities, which is out of the observation ability of LAMOST. The large number of YSG candidates is as expected since the photometric colors can not distinguish YSGs with (a large number of) foreground contaminants.

\begin{table}
\begin{center}
\caption[]{Number of supergiant candidates of M31/M33 in LAMOST DR10 selected based on {\it UBVRI} photometry. Amongst the 924 YSG candidates of M31 , 111 are selected based on photometry taken from \citet{Magnier1992} because those targets are too bright for LGGS. }\label{tab:targets}
\begin{tabular}{cccc}
    \hline\noalign{\smallskip}
Galaxy &  BSG candidates & YSG candidates & RSG candidates \\
M31 & 108 & 908 & 6 \\
M33 & 80 & 729 & 5 \\
  \noalign{\smallskip}\hline
\end{tabular}
\end{center}
\end{table}

\begin{table}
\begin{center}
\caption[]{Number of LAMOST supergiant candidates of M31/M33 in LAMOST DR10 selected based on {\it Gaia} photometry.}\label{tab:targets_gaia}
\begin{tabular}{cccc}
    \hline\noalign{\smallskip}
Galaxy &  BSG candidates & YSG candidates & RSG candidates \\
M31 & 27 & 2230 & 1 \\
M33 & 7 & 758 & 1 \\
  \noalign{\smallskip}\hline
\end{tabular}
\end{center}
\end{table}

\section{Supergiants identification}\label{sect:identify}
Here we introduce our methods on separating M31/M33 supergiants from foreground dwarfs. As mentioned earlier, separating M31/M33 supergiants from foreground dwarfs based solely on their positions on the color-magnitude diagram or intrinsic colors, especially for YSGs, is quite difficult \citep{Drout2009}. Fortunately, M31/M33's kinematics allow us to overcome this problem (\citealt{Massey2009,Massey2016,Drout2009,Drout2012}). The two galaxies have a large negative systemic radial velocities (about $-300$\,km\,s$^{-1}$ for M31 and $-200$\,km\,s$^{-1}$ for M33) and are both rotating systems. Following \citet{Drout2009,Drout2012}, one can predict radial velocity (RV) of M31/M33 members according to their positions ($X$ and $R$):

\begin{equation}\label{eq:vexp_31}
  V_{\rm exp}^{\rm M31} = -295 + 241.5(X/R),
\end{equation}
and
\begin{equation}\label{eq:vexp_33}
  V_{\rm exp}^{\rm M33} = -182 - 81.9(X/R),
\end{equation}
where $X$ is the distance along the M31/M33's semimajor axis, and $R$ is the radial distance of the object within the plane of M31/M33 (\citealt{Drout2009,Drout2012}). By comparing the expected radial velocities, $V_{\rm exp}$, to the observed ones, the M31/M33 members can be well selected out from the contaminations.

\subsection{Observed Radial Velocities from LAMOST LRS}\label{subsect:obs_rv}
\begin{figure}[htbp!]
   \centering
   \includegraphics[width=\textwidth, angle=0]{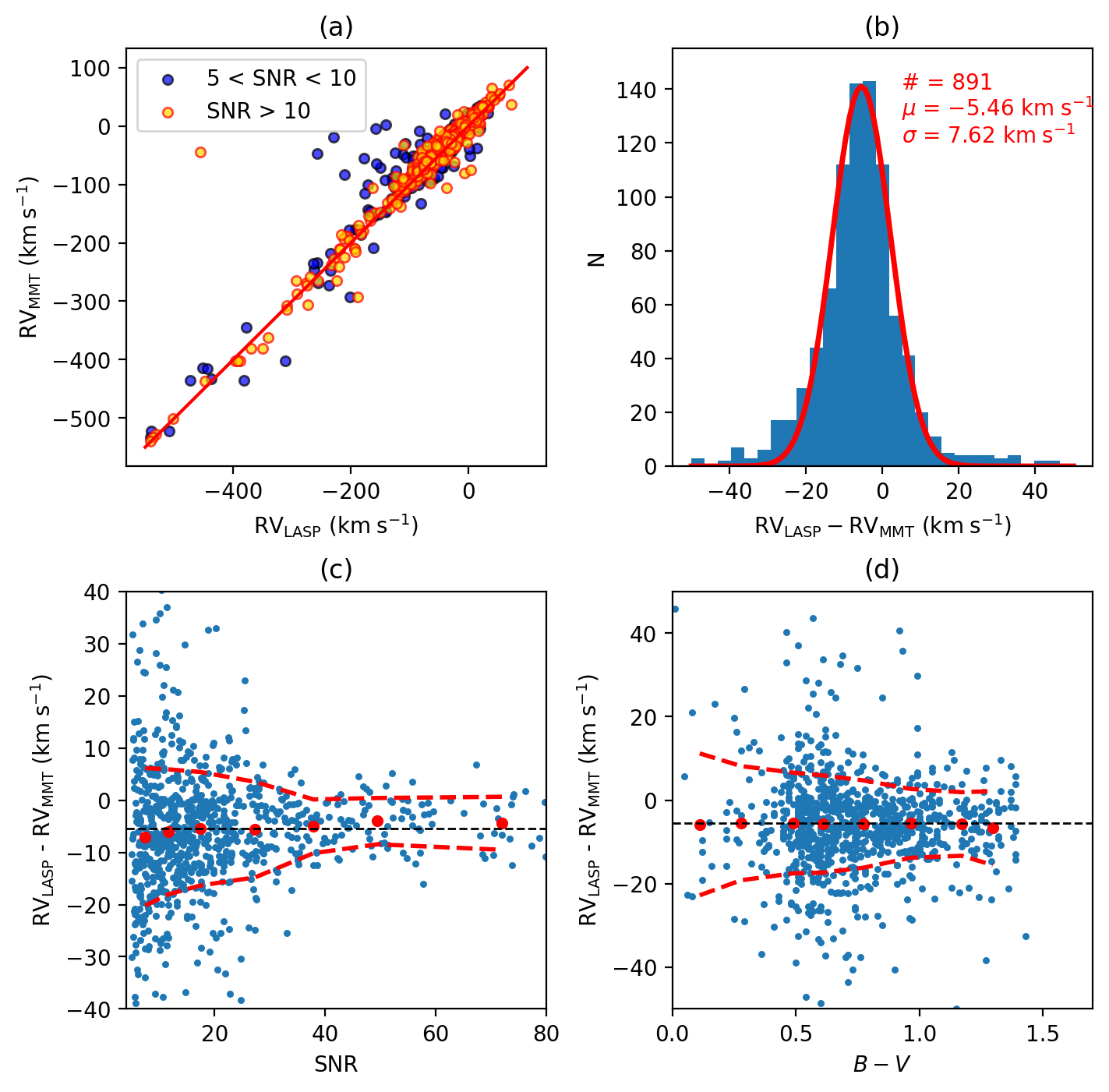}
   \caption{Comparison between observed RVs of 891 objects derived from LASP and MMT. Panel (a) shows the distribution of observed RVs, with the red line indicating the 1:1 line. In panel (b), the histogram illustrates the differences between LASP and MMT observed RVs, with the red line representing a Gaussian fit. Median and standard deviation from the Gaussian fit are also marked in the top-left corner. Panels (c) and (d) show the RV differences in relation to SNR and  $B - V$ color, respectively, with the black dashed line representing the median difference. The red dots in panels (c) and (d) represent the mean difference value of each SNR and $B - V$ bin, with the red dashed lines indicating the corresponding 1$\sigma$ region.}
   \label{fig:m31_m33_rvobs}
   \end{figure}

3936 spectra of a total of 3280 supergiant candidates in M31 and 1580 ones in M33 have been obtained from the LAMOST DR10 LRS. For sources with mutiple observations, we take the one with highest SNR. The RV measurements, as well as the stellar atmospheric parameters, are derived by the LAMOST stellar parameter pipeline (LASP; \citealt{Luo2015}). Through comparisons with RV measurements obtained from high-resolution spectroscopy and RV standard stars (\citealt{Luo2015,Gao2015,Huang2018rv,Li2023}), the zero offset of LAMOST LRS RV is found to be around $-5$\,km\,s$^{-1}$, and the typical precision is also around $-5$\,km\,s$^{-1}$. 
As an independent check, the LAMOST LRS RVs are compared to those derived from MMT spectra by Massey's group (\citealt{Massey2009,Drout2009}). Overall, they exhibit excellent consistency, with no significant trends detected along both SNR and stellar color $B - V$ (see Fig.~\ref{fig:m31_m33_rvobs}). The scatter is only $6.79$ km\,s$^{-1}$ for stars with SNR\,$>10$ and $11.46$ km\,s$^{-1}$ even at $5 <$\,SNR\,$<10$. The overall median offset is $-5.46$\,km\,s$^{-1}$, in excellent consistent with that found by aforementioned studies. An offset of 5\,km\,s$^{-1}$ was therefore added to RVs derived from LAMOST LRS spectra.

\subsection{Identifying Supergiants}
\label{subsect:identify_31}
\begin{figure}[htbp!]
  \begin{minipage}[t]{0.497\linewidth}
  \centering
   \includegraphics[width=75mm]{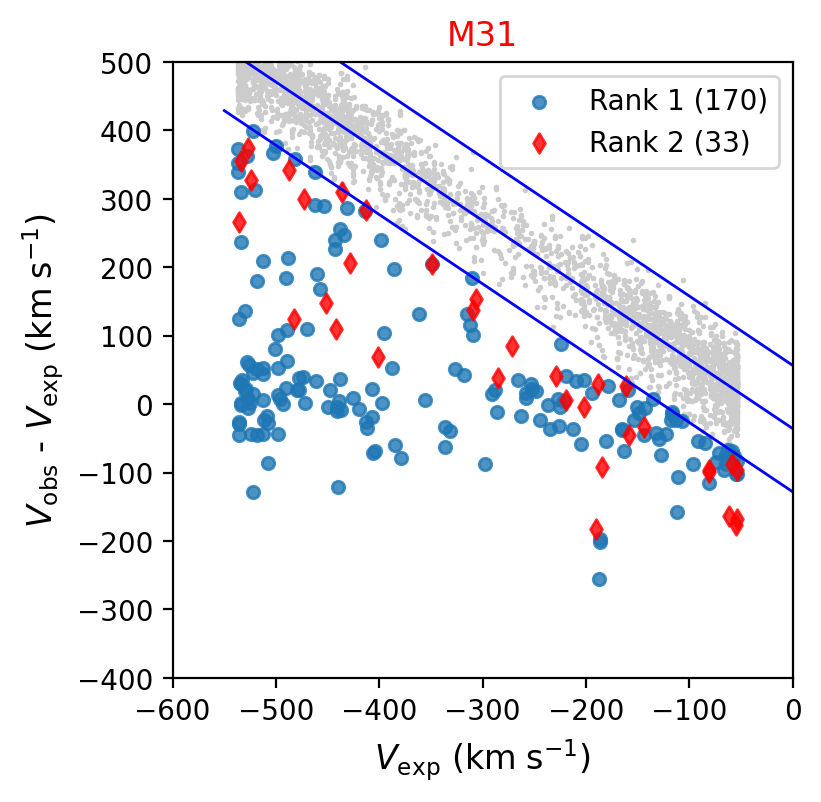}
  \end{minipage}
  \begin{minipage}[t]{0.497\textwidth}
  \centering
  \includegraphics[width=75mm]{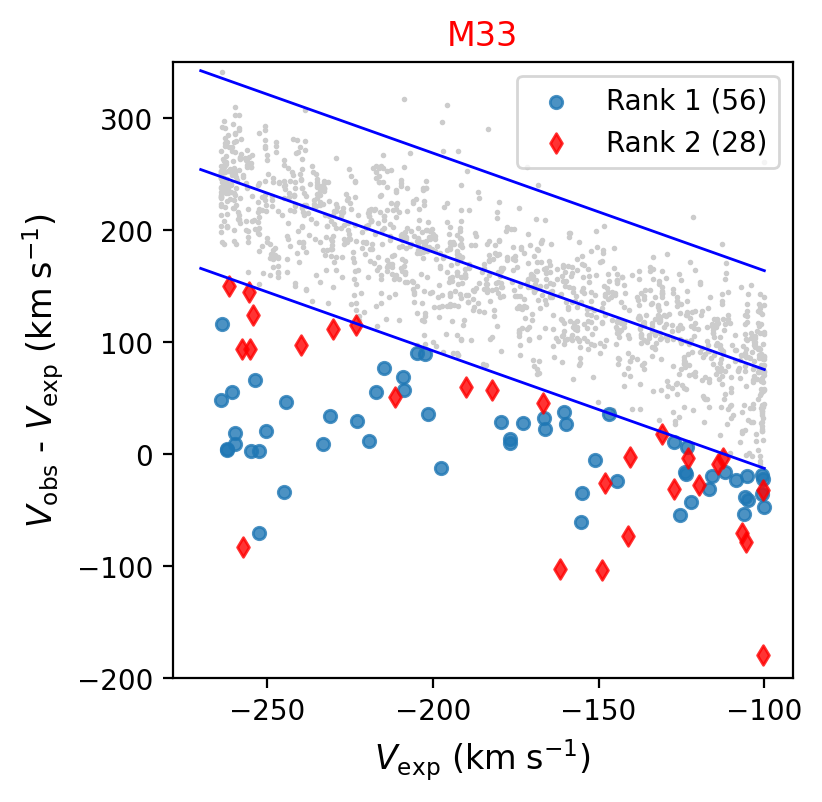}
  \end{minipage}
  \caption{ RV distributions of M31 (left) and M33 (right) supergiant candidates. The blue lines represent a linear fit to the diagonal branch, along with the $\pm3\sigma$ confidence region. Candidates that passed both the RV and {\it Gaia} astrometry selections are labeled as `Rank1' (blue dots), while those only passed RV screening are designated as `Rank2' (red diamonds). }
  \label{fig:rv_rank1_2}
\end{figure}

Now armed with the observed RVs of our supergiant candidates, foreground contaminations could be eliminated by comparing expected RVs predicted from Eqs.~\ref{eq:vexp_31} and~\ref{eq:vexp_33} to the observed ones. 
In the left panel of Fig.~\ref{fig:rv_rank1_2}, two branches are clearly visible: the one with nearly constant value of zero represents the member stars of M31, while the significant diagonal branch indicates the foreground field stars of the Milky Way. We note the two branches gradually overlap each other with increasing $X$. This suggests that candidates found in north-eastern corner may still suffer moderate contaminations from the foreground stars. 
A linear fit was performed to the diagonal branch (represented by the middle solid blue line in the left panel of Fig.~\ref{fig:rv_rank1_2}), and a Gaussian function was applied to the fitting residuals to determine its scatter $\sigma$. 210 objects falling below the diagonal branch (represented by the best-fit) minus $3\sigma$ are considered as M31 supergiant candidates. Among them, 185 were selected from LGGS or \cite{Magnier1992} and 25 were selected from {\it Gaia} DR3 photometry. 39 of these 210 candidates have been observed at least two times by LAMOST with SNR $>$ 5. For these, we compare the RV values from different observations and remove 4 candidates with variation (defined as the difference between maximum and minimum) larger than 50 km\,s$^{-1}$. This number (4/39) also indicates that the foreground contamination among the M31 supergiant candidates identified with only one observation is no more than 10\%.
The spectra of the remaining 206 candidates are then visually inspected and 7 of them are discarded due to their bad qualities.

\begin{figure}
   \centering
   \includegraphics[width=\textwidth, angle=0]{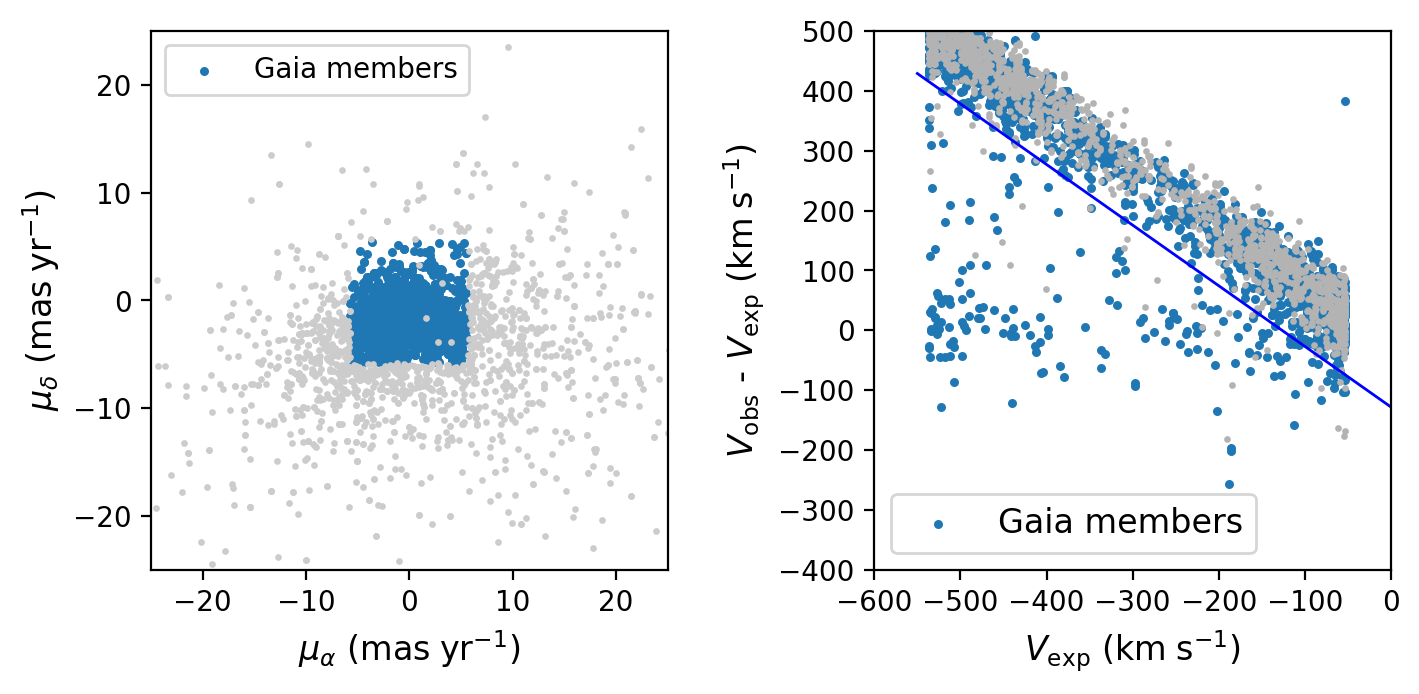}
   \caption{Left: Proper motions of M31 supergiant candidates. The gray dots represent those foregrounds defined by using {\it Gaia} astromtry selection, with parallaxes or proper motions falling outside the region that contains 99.5\% of the comparison sample. M31 members defined this way are denoted as blue dots. Right: RV distribution of M31 supergiant candidates selected astrometrically. Foregrounds defined by using {\it Gaia} astrometry selection are denoted as gray dots. The blue line represents the area minus 3$\sigma$ away from the diagonal branch.}
   \label{fig:m31_gaia}
   \end{figure}

In principle, astrometric measurements from {\it Gaia} provide another way to distinguish supergiant members of M31 from foreground field stars. Doing so, the numerous RSG candidates of M31 identified by \cite{Massey2021} based on near infrared (NIR) colors, as well as the {\it Gaia} DR2 astrometry, are utilized as a comparison sample. This comparison sample provided a reference for the distribution of parallax and proper motions of young disk-like objects (similar to those in this study) in M31. In total, near 4500 RSG candidates in M31 are found with astrometric information from {\it Gaia} DR3, with an average parallax of $0.06 \pm 0.74$ mas, and mean proper motions in right ascension and declination of $-0.15 \pm 1.19$ and $0.03 \pm 1.36$ mas$\,{\rm yr^{-1}}$, respectively. Following a similar methodology in \cite{Massey2021}, we identify foreground stars among the photometrically selected supergiant candidates (see Tables~\ref{tab:targets} and~\ref{tab:targets_gaia}), with parallax or proper motions falling outside the region that contains 99.5\% of the comparison sample. As clearly shown in Fig.~\ref{fig:m31_gaia}, the foreground stars defined this way largely (97.5\%) distribute in the diagonal branch of the RV differences diagram, with only a few falling below the diagonal branch minus $3\sigma$. Similar case is found for M33. This result confirms the effectiveness of the RV measurements on separating M31/M33 members from foreground contaminations.

Therefore, we combine both RV and {\it Gaia} astrometry criteria to select supergiant members. Candidates exhibiting RV differences below the diagonal branch minus $3\sigma$, along with parallax and proper motions within the region containing 99.5\% of the comparison sample, are classified as `Rank1' candidates. These candidates have successfully passed through two independent selection criteria, indicating a high level of credibility as true supergiant members of M31/M33. On the other hand, candidates with RV distributions similar to those of `Rank1' candidates, but exhibiting a significant deviation of parallax or proper motions from the mean values of the comparison sample, are classified as `Rank2' candidates.
In total, there are 168 `Rank1' and 31 `Rank2' supergiant candidates of M31, represented as blue dots and red diamonds in the left panel of Fig.~\ref{fig:rv_rank1_2}, respectively. 

Employing the same methodology as M31, we conducted a selection of supergiant candidates of M33. 
We began by comparing the observed and the expected RVs of the photometrically selected supergiant candidates of M33, deriving the best linear regression for objects within the diagonal line and the corresponding scatter $\sigma$, as shown in the right panel of Fig.~\ref{fig:rv_rank1_2}. 85 objects falling below the diagonal branch minus 3$\sigma$ were selected as M33 supergiant candidates. Of the 85 candidates, 17 have been observed more than once by LAMOST with SNR $>$ 5, and none of them shows RV variation larger than 50 \,km\,s$^{-1}$. Then after a visual inspection of their spectra, one candidate is excluded due to poor quality.
Subsequently, we utilized approximately 2100 M33 RSG candidates from \cite{Massey2021} with available {\it Gaia} DR3 astrometric measurements as a comparison sample to double-check foreground contaminations. 
Following the same procedure used for M31, we identified 56 `Rank1' and 28 `Rank2' M33 supergiant candidates. These are represented by blue dots and red diamonds, respectively, in the right panel of Fig.~\ref{fig:rv_rank1_2}.

\subsection{Membership Properties and Re-examination}\label{subsect:re-examine}

\begin{figure}
\centering
\includegraphics[width=\textwidth, angle=0]{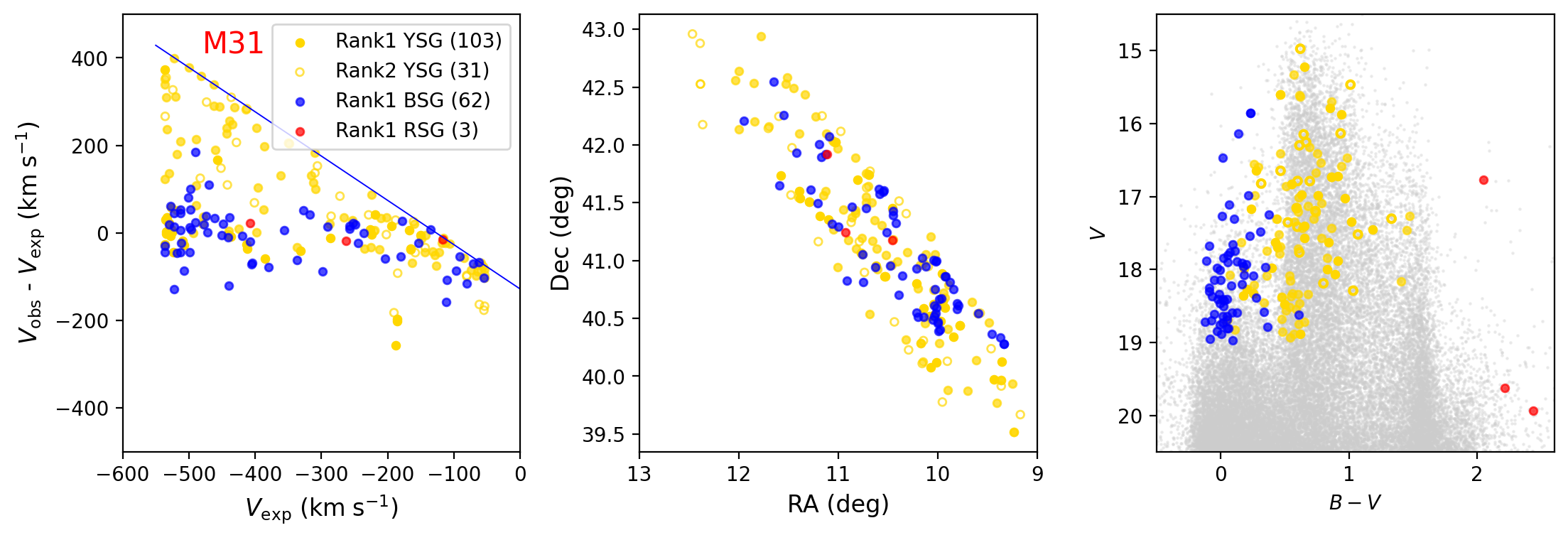}
\caption{Left panel: The RV distribution of different types of supergiant candidates of M31. `Rank1' YSG, BSG, and RSG candidates are represented by yellow, blue, and red solid circles respectively, while `Rank2' YSG candidates are marked as open yellow circles.
Middle panel: The spatial distribution of different types of supergiant candidates of M31.
Right panel: The color-magnitude $B - V$ versus $V$ diagram of different types of supergiant candidates of M31. Background gray dots represent the LGGS objects in M31 sky region.}
\label{fig:m31_properties}
\end{figure}

\begin{figure}
\centering
\includegraphics[width=\textwidth, angle=0]{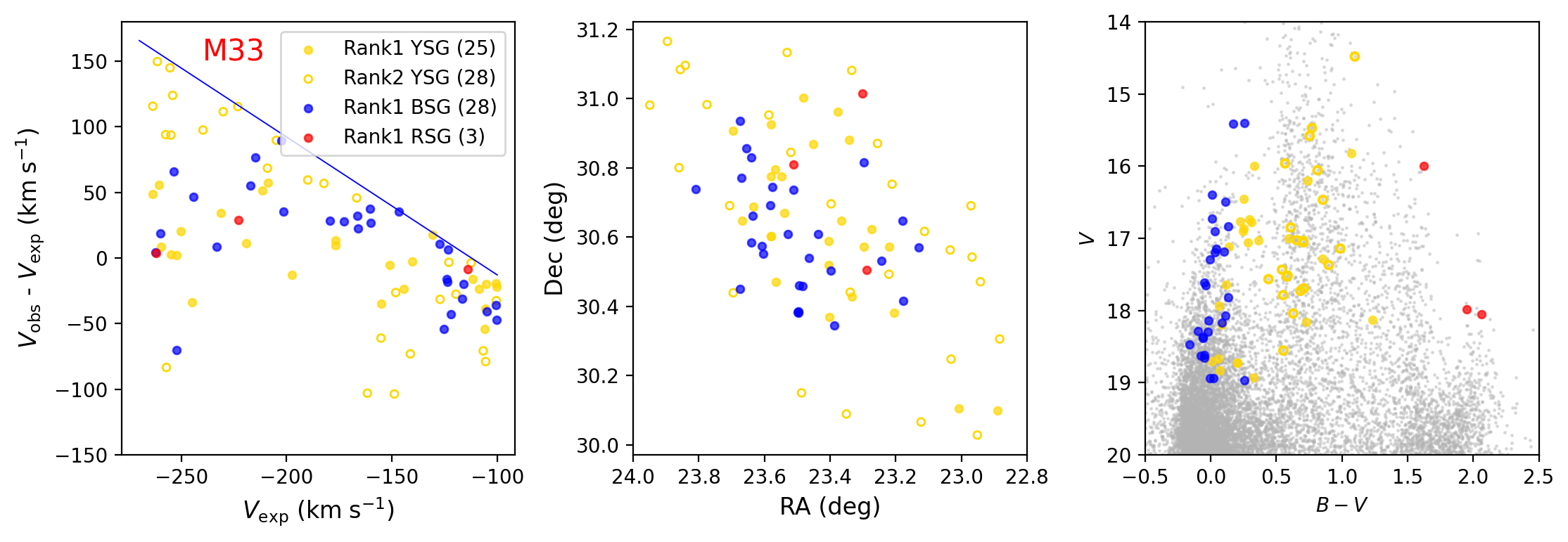}
\caption{Similar to Fig.~\ref{fig:m31_properties}, but for M33 supergiant candidates.}
\label{fig:m33_properties}
\end{figure}

In Figs.~\ref{fig:m31_properties} and~\ref{fig:m33_properties}, the RV and spatial distribution, as well as the color-magnitude diagram, of different types of supergiant candidates in M31/M33 are presented. Note that the color-magnitude diagram in the right panel does not contain candidates without {\it B} and {\it V} photometry. 
In the left panel, it is evident that majority of BSGs and RSGs locate near the region where RV differences are approximately zero, indicating excellent agreement between the observed RVs and the predicted RVs. In contrast, a fraction of YSGs' observed RVs deviate from the predicted ones. Additionally, YSG candidates are more widely distributed in the disk of M31/M33 compared to BSGs and RSGs.
Furthermore, in the color-magnitude diagrams, the bluest YSGs in both M31 and M33 extend into the BSG region. This extension is consistent with our loosening of the original color selection to include stars as blue as $B - V = 0$, similar to the criteria set by \cite{Drout2009}.

Considering that the RVs of quite a lot of YSG candidates deviating the predicted ones, it is essential to re-examine the spectra of those YSGs, so as to validate their membership of M31.
The presence of the O$\mathrm{I}$ $\lambda$7774 triplet serves as a robust indicator for identifying YSGs. O$\mathrm{I}$ $\lambda$7774 triplet is known for its strong luminosity effect in F-type supergiants due to non-LTE effects \citep{Osmer1972}. The strengths of O$\mathrm{I}$ $\lambda$7774 can be accentuated by spherical mass outflows of typical supergiants \citep{Przybilla2000}. Candidates exhibiting this feature are considered firmed YSGs \citep{Drout2009}. Unfortunately, this feature is hard to be detected in LAMOST spectra due to low SNR and sky emission subtraction issues. Finally, only nine spectra of YSGs show significant O$\mathrm{I}$ $\lambda$7774 absorption lines.
In future, follow-up high-quality spectroscopy are required to confirm the memberships of those YSG candidates, especially those not siting around the zero line region (observed minus predicted ones in RV distribution diagram).

\subsection{Comparison With Previous Studies}
\label{subsect:compare_with_massey}
As mentioned in Secion~\ref{sect:intro}, Massey's group has conducted a series of studies to identify supergiant members. Therefore, we cross-matched our candidates with the catalogs published from \cite{Massey2016} and \cite{Massey2021}.
In total, we found that among the 168 `Rank1' candidates of M31, 87 candidates had been studied by Massey's group and were already in the catalog from \cite{Massey2016}, with an additional 2 RSG candidates included in the catalog by \cite{Massey2021} through NIR photometry. Among those 87 candidates, 63 were identified as members, possible members, or unknown, while the other 24 objects were classified as foregrounds by \cite{Massey2016}. These 63 candidates, together with the 2 RSG candidates identified through NIR, are referred to `Massey members'. On the other hand, for those 24 objects classified as foregrounds by \cite{Massey2016}, we designate them as `Massey non-members'. Additionally, 6 `Massey members' and 7 `Massey non-members' are found among our 31 `Rank2' M31 supergiant candidates. For M33, the 56 `Rank1' supergiant candidates include 36 `Massey members' and 1 `Massey non-member'. Among the 28 `Rank2' candidates of M33, there are 5 `Massey members' and 12 `Massey non-members'. We note that most of these `Massey non-members' are YSGs.

\begin{figure}[h]
  \begin{minipage}[t]{0.497\linewidth}
  \centering
   \includegraphics[width=75mm]{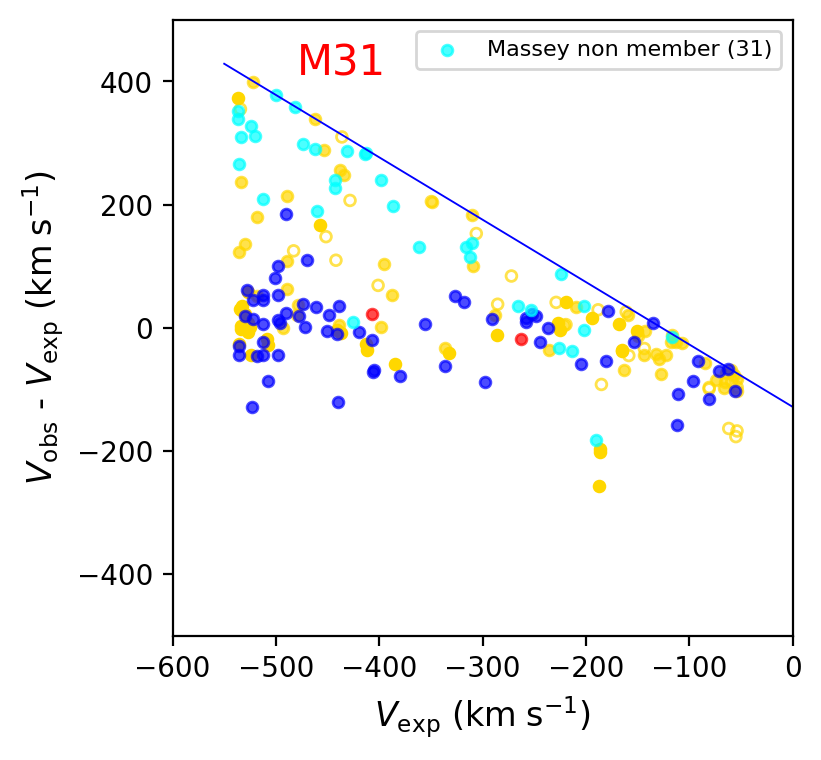}
  \end{minipage}
  \begin{minipage}[t]{0.497\textwidth}
  \centering
  \includegraphics[width=75mm]{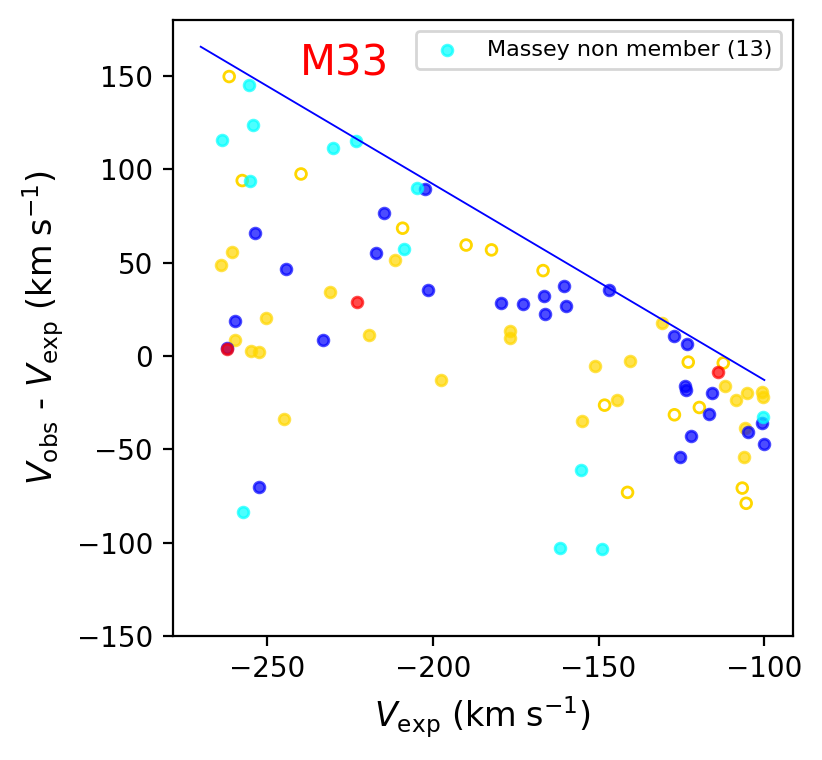}
  \end{minipage}
  \caption{ Similar to the left panels of Figs.~\ref{fig:m31_properties} and~\ref{fig:m33_properties}, but with `Massey non-members' marked as cyan dots. Left panel is for M31 and right panel is for M33. }
  \label{fig:m31_compare}
\end{figure}

We checked the RV distribution of those `Massey non members' in Fig.~\ref{fig:m31_compare}. The RV differences of most of them are very close to the diagonal branch minus $3\sigma$. Considering none of them are confirmed with the O$\mathrm{I}$ $\lambda$7774 triplet, their memberships of M31/M33 are required to be explored by further observations as suggested earlier.

\section{Result and discussion}
\label{sect:result}
We present the supergiant candidates identified in this work in Tables~\ref{tab:new_sg_m31} and~\ref{tab:new_sg_m33}. A more detailed version of the catalogues, including RA, DEC, and LAMOST observation times with SNR $>$ 5, is available in electronic form in the online version of this manuscript\footnote{Online version has been published in https://www.scidb.cn/en/s/F3Yn2q}.

\subsection{Physical Properties and HR-Diagram}\label{subsect:hrd}

In order to examine the distribution of our supergiant candidates of M31/M33 within the Hertzsprung-Russell Diagram (HRD) and check their consistency with current stellar evolutionary tracks, it is essential to derive their effective temperatures and bolometric luminosities.

For candidates with {\it B} and {\it V} photometry available in M31, we apply a constant reddening correction of $E(B - V) = 0.13$. This is the median value found for early-type stars in M31 by \cite{Massey2007a} and is widely employed in the analysis of M31's supergiants. For stars with {\it Gaia} photometry, the reddening coefficient is then taken from \cite{Huang2021}. As our candidates, mostly BSGs and YSGs, cover a wide range of dereddened colors $(B-V)_{0}$ from $-0.25$ to $2.31$, which cover a wide range of effective temperatures. \cite{Drout2009} provided transformations using the ``Atlas9" model \citep{Kurucz1992} for objects with $0.03$ $\leq$ $(B-V)_{0}$ $\leq$ $1.26$. A significant fraction of our candidates fall outside this range. Consequently, we opt to employ data from YBC database \citep{Chen2019} to derive effective temperature-color relations and bolometric correction (BC) with a wide application range of color. The BC values in different temperature ranges in the YBC database are the optimized results provided after comparing the results derived from various models (more detailed in \citealt{Chen2019}). Additionally, the YBC database is allowed to choose different stellar mass and metallicity. 
By restricting the initial mass within the typical range for supergiants and fixing the metallicity as 2 $Z_{\odot}$, we obtain the corresponding transformation from dereddened colors $(B-V)_{0}$ to effective temperatures, encompassing a broader color range $-0.37$ $\leq$ $(B-V)_{0}$ $\leq$ $1.53$, as well as the BCs. Note that 3 RSGs are not included in the transforming process due to their extremely high $B-V$ values, and therefore not included in the HRD below.
   
The relationship between $(B-V)_{0}$ and log $T_{\rm eff}$ for supergiant candidates in M31 is as follows: 
\begin{equation}\label{eq:teff_transfer_31}
{\rm log}\ T_{\rm eff} = \begin{cases}
    3.94 - 0.8644\,(B-V)_{0} + 4.424\,(B-V)_{0}^2, & (B-V)_{0} \leq  0.03, \\
    3.924 - 0.3812\,(B-V)_{0} + 0.3607\,(B-V)_{0}^2 - 0.1722\,(B-V)_{0}^3, & (B-V)_{0} > 0.03.
    \end{cases}
\end{equation}
The corresponding BCs are as follows: 
\begin{equation}\label{eq:bc_31}
    {\rm BC} = \begin{cases}
    -221.1 + 114.8\,{\rm log}\ T_{\rm eff} - 14.89\,({\rm log}\  T_{\rm eff})^2, & {\rm log}\ T_{\rm eff} \leq 4.1,\\
    21.98 - 5.544\,{\rm log}\ T_{\rm eff}, & {\rm log}\ T_{\rm eff} > 4.1.
    \end{cases}
\end{equation}
For candidates with only available $Gaia$ photometry, we establish similar transformation from $(G_{\rm BP}-G_{\rm RP})_{0}$ to log $T_{\rm eff}$ and the corresponding BC values for $(G_{\rm BP}-G_{\rm RP})_{0}$ ranging from $-0.1$ to $1.24$, based on data from YBC database:
\begin{equation}\label{eq:teff_transfer_31_gaia}
\begin{gathered}
    {\rm log}\ T_{\rm eff} = 3.953 - 0.3905\,(G_{\rm BP}-G_{\rm RP})_{0} + 0.3081\,(G_{\rm BP}-G_{\rm RP})_{0}^2 - 0.1263\,(G_{\rm BP}-G_{\rm RP})_{0}^3,\\
    {\rm BC} = - 179.4 + 93.89\,{\rm log}\ T_{\rm eff} - 12.27\,({\rm log}\  T_{\rm eff})^2.
\end{gathered}
\end{equation}

For M33 candidates, we apply a constant reddening correction of $E(B - V) = 0.12$, and compute the transformation relationship based on the YBC database. We set the metallicity as 0.6 $Z_{\odot}$, in line with \cite{Drout2012}.
For candidates with {\it B} and {\it V} photometry available, the relationship between $(B-V)_{0}$ and log $T_{\rm eff}$ are as follows:
\begin{equation}\label{eq:teff_transfer_33}
    {\rm log}\ T_{\rm eff} = \begin{cases}
    3.942 - 1.159\,(B-V)_{0} + 5.187\,(B-V)_{0}^2 + 8.167\,(B-V)_{0}^3, & (B-V)_{0} \leq 0.06,\\
    3.907 - 0.318\,(B-V)_{0} + 0.1899\,(B-V)_{0}^2 - 0.06115\,(B-V)_{0}^3, & (B-V)_{0} > 0.06.
    \end{cases}
\end{equation}
The BCs are as follows: 
\begin{equation}\label{eq:bc_33}
    {\rm BC} = \begin{cases}
    -268.4 + 139.6\,{\rm log}\ T_{\rm eff} - 18.15\,({\rm log}\ T_{\rm eff})^2, & {\rm log}\ T_{\rm eff} \leq 4.0,\\
    3.419 + 6.296\,{\rm log}\ T_{\rm eff} - 1.378\,({\rm log}\ T_{\rm eff})^2, & {\rm log}\ T_{\rm eff} > 4.0.
    \end{cases}
\end{equation}
For candidates of M33 selected based on {\it Gaia}, the transformation and BC for objects with $(G_{\rm BP}-G_{\rm RP})_{0}$ ranging from 0.2 to 1.4 are as follows:
\begin{equation}\label{eq:teff_transfer_33_gaia}
\begin{gathered}
    {\rm log}\ T_{\rm eff} = 3.933 - 0.3107\,(G_{\rm BP}-G_{\rm RP})_{0} + 0.1895\,(G_{\rm BP}-G_{\rm RP})_{0}^2 -0.08074\,(G_{\rm BP}-G_{\rm RP})_{0}^3,\\
    {\rm BC} = - 162.2 + 84.91\,{\rm log}\ T_{\rm eff} - 11.1\,({\rm log}\ T_{\rm eff})^2.
\end{gathered}
\end{equation}

Applying distance modulus of 24.40 and 24.60 for M31 and M33, respectively, as taken from \cite{Van2000}, we derive the bolometric luminosities of all M31/M33's `Rank1' and `Rank2'  supergiants. 

\begin{figure}[htbp!]
  \begin{minipage}[t]{0.497\linewidth}
  \centering
   \includegraphics[width=75mm]{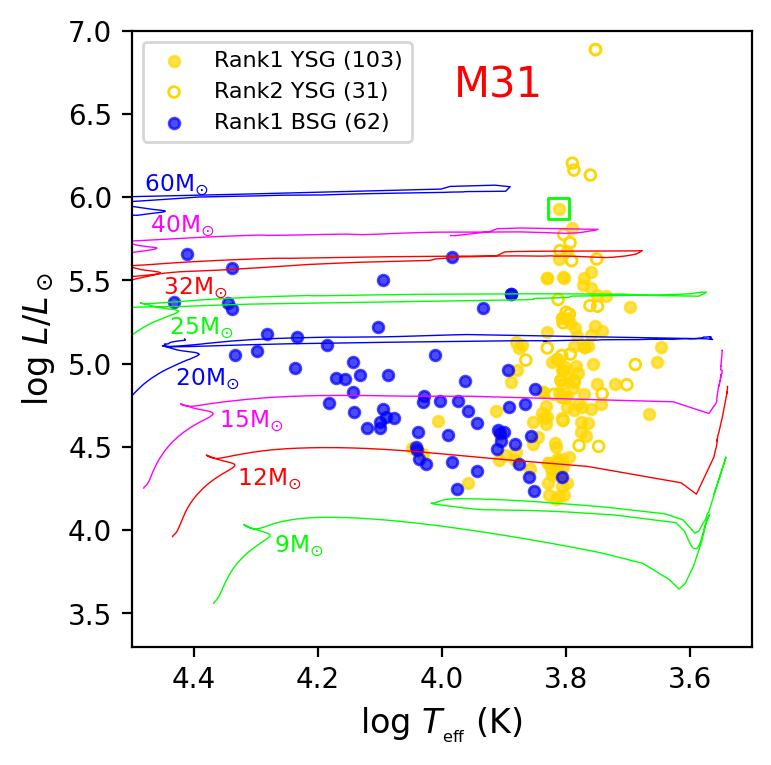}
  \end{minipage}
  \begin{minipage}[t]{0.497\textwidth}
  \centering
   \includegraphics[width=75mm]{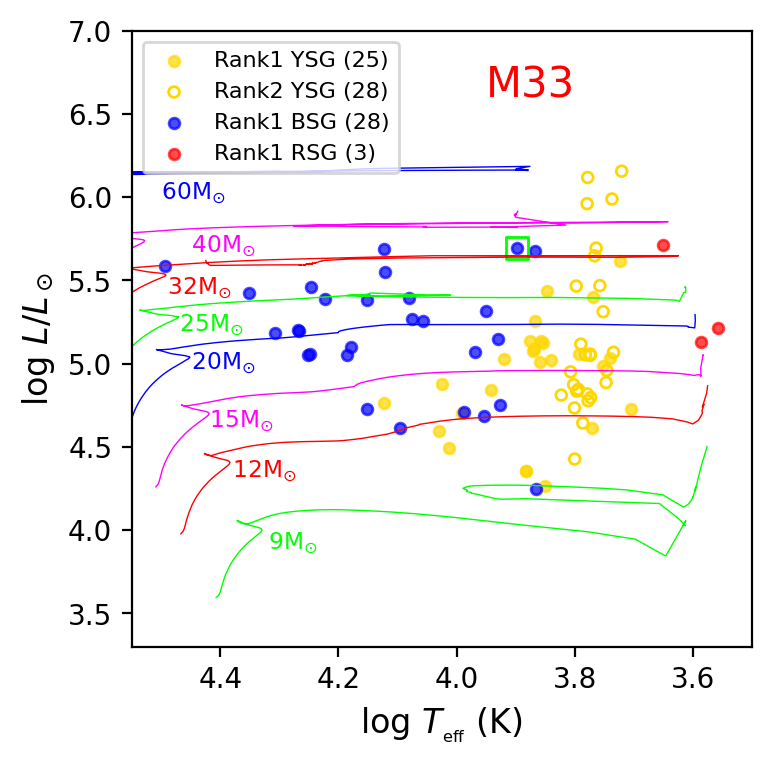}
  \end{minipage}
  \caption{\small HRDs for `Rank1' and `Rank2' supergiant candidates of M31 (left) and M33 (right). The overplotted lines with different colors show the Geneva evolutionary tracks for 2 times solar metallicity for M31 and 0.6 times solar metallicity for M33, with an initial rotation speed of 40\% of the breakup speed. The green boxes enclose the most massive `Rank1' candidates of M31 and M33.}
  \label{fig:m31_m33_HRD}
\end{figure}

In Fig.~\ref{fig:m31_m33_HRD}, we present the locations of supergiant candidates of M31/M33 in the HRD. Geneva evolutionary tracks for 2 and 0.6 $Z_{\odot}$, respectively, with an initial rotation speed of 40\% of the breakup speed, are overplotted \citep{Yusof2022}. 
Notably, in the left panel of Fig.~\ref{fig:m31_m33_HRD}, there are 4 `Rank2' YSG candidates of M31 with extremly high luminosities, inconsistent with the evolutionary tracks. A similar scenario is observed in the right panel of Fig.~\ref{fig:m31_m33_HRD}, where the most luminous 4 `Rank2' YSGs deviate from the evolutionary tracks. Considering their `Rank2' properties, without the O$\mathrm{I}$ $\lambda$7774 triplet confirmed in their spectra, we doubt their status as genuine YSGs, and further confirmation is required.
The locations of other supergiant candidates exhibit a excellent agreement with the evolutionary tracks.

Among our `Rank1' candidates of M31, the most massive one is LAMOST J0043+4124, highlighted by a green box in the left panel of Fig.~\ref{fig:m31_m33_HRD}, with a mass of over 40 $M_{\odot}$. This object has the brightest apparent {\it G} magnitude among our `Rank 1' candidates and locates in the region with expected RVs greater than $-100$ km\,s$^{-1}$. In this region, the M31 members and the foregrounds may not be well separated by RVs \citep{Hartmann1997}. 
However, considering that it passes both the RV and Gaia astrometry criteria successfully, we have chosen to retain it as a YSG candidate.
In the case of M33, the most massive one among the new `Rank1' candidates is a BSG, J0134+3044. Highlighted by a green box in the right panel of Fig.~\ref{fig:m31_m33_HRD}, it is estimated to be more than $\sim$ 32 $M_{\odot}$ based on evolutionary tracks. And this object has been analyzed by \cite{Liu2022} through spectroscopic methods. The effective temperature and bolometric luminosity derived by \cite{Liu2022} are in agreement with our estimations in this work, indicating the reliability of our newly derived transformation relations.

\subsection{Spatial distribution and possible substructures}\label{subsect:structure}
\begin{figure}[htbp!]
   \centering
   \includegraphics[width=\textwidth, angle=0]{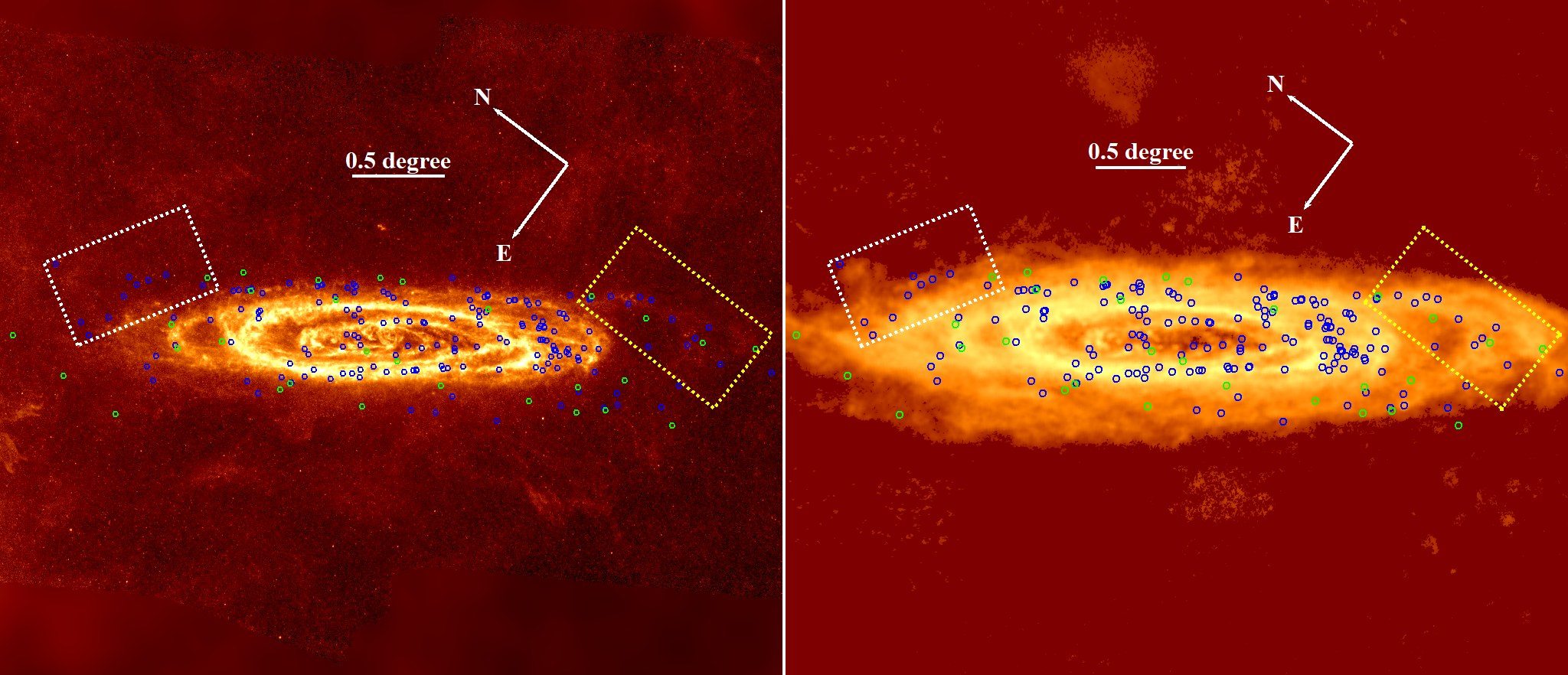}
   \caption{Left: Spatial distribution of supergiant candidates of M31 based on {\it Herschel} SPIRE 250 $\mu$m image \citep{Fritz2012}. The blue circles mark the locations of `Rank1' supergiant candidates, while the green circles represent the positions of `Rank2'. The white and yellow dashed boxes highlight substructures in the northeastern and southwestern corners, respectively. Spatial scale and compass are shown in upper. Right: Spatial distribution of supergiant candidates of M31 based on HI 21 cm map \citep{Braun2009}.}
   \label{fig:m31_structure}
   \end{figure}

In Fig.~\ref{fig:m31_structure}, we present the spatial distribution of our `Rank1' and `Rank2' supergiant candidates in {\it Herschel} SPIRE 250 $\mu$m \citep{Fritz2012} and HI 21 cm \citep{Braun2009} images of M31. In the left panel, the majority of the supergiant candidates fall along the CO ring, which corresponds to the high star formation region, in line with our expectations. Additionally, several supergiant candidates are found outside the CO ring, including two extended substructures of M31's disk. One is in the southwestern corner, which has already been validated as a part of M31 (\citealt{Fritz2012,Braun2009}). This substructure is highlighted by a yellow box in Fig.~\ref{fig:m31_structure}, and the LBV found by \cite{Huang2019} is situated in this substructure. The other substructure is located in the northeast corner and includes 11 candidates, highlighted by a white box in Fig.~\ref{fig:m31_structure}, commented as `NE' in Table~\ref{tab:new_sg_m31}. Among these candidates, 10 are YSGs (9 `Rank1' and 1 `Rank2') and 1 is a BSG (`Rank1'). In the right panel, these candidates are found near the gas ring. This region faces significant contamination from foreground Galactic emission, suggesting the observed gas may predominantly originate from the Milky Way, not M31 itself.
\cite{Fritz2012} analyzed this region using {\it Herschel} far-infrared data, correcting for the foreground dust component, but could not conclusively determine whether the substructure belongs to the Galactic foreground or M31. If follow-up observations confirm that the 11 supergiant candidates are members of M31, it would provide strong evidence for the ownership of this substructure. This will be a significant component of our future work.
   
\begin{figure}[htbp!]
   \centering
   \includegraphics[width=\textwidth, angle=0]{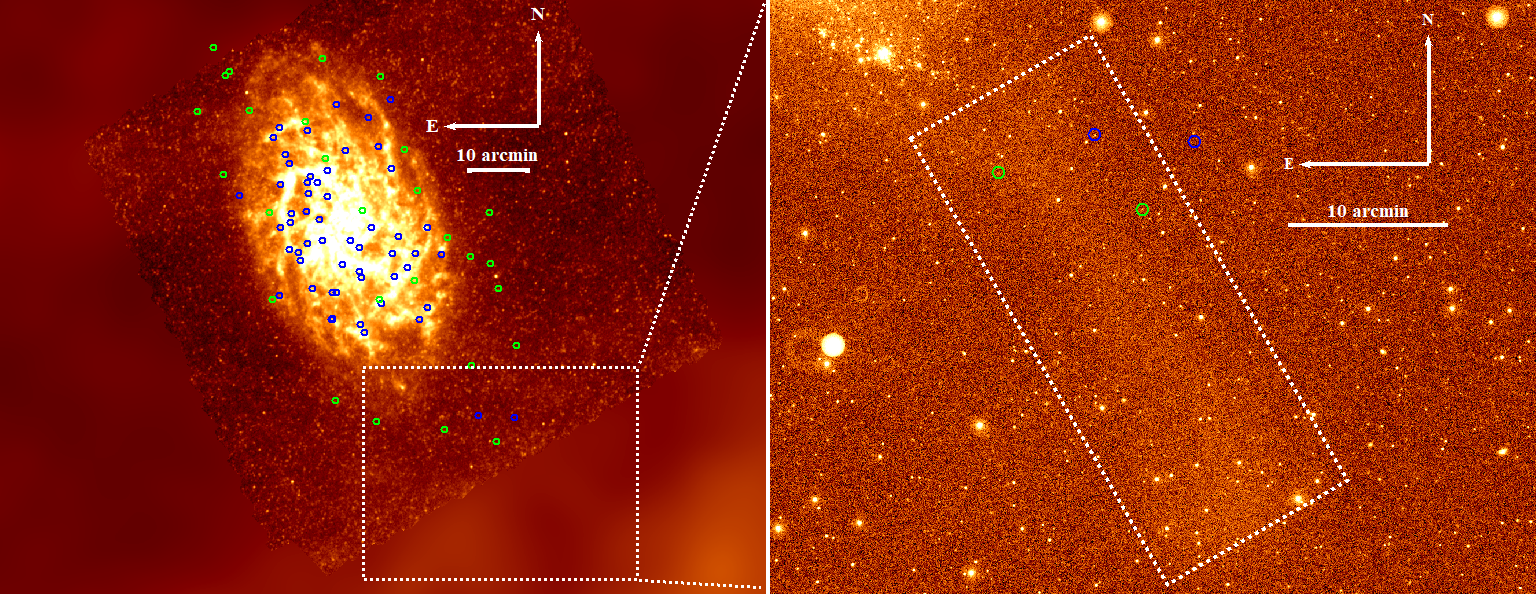}
       \caption{Left: Spatial distribution of supergiant candidates of M33 {\it Herschel} SPIRE 250 $\mu$m image. The blue circles mark the locations of the `Rank1' supergiant candidates, and the green circles represent those of `Rank2'. The spatial scale and compass are displayed in the top-right. Right: Image of the southwestern corner of M33 in {\it Galex} NUV band. The white dashed box highlights an extended substructure from M33's disk, where three supergiant candidates are distributed.}
   \label{fig:m33_structure}
   \end{figure}

In the left panel of Fig.~\ref{fig:m33_structure}, we present the spatial distribution of our supergiant candidates in {\it Herschel} SPIRE 250 $\mu$m image of M33. Similar to M31, the majority of supergiant candidates are distributed along the CO ring. Additionally, 4 YSG candidates (2 `Rank1' and 2 `Rank2') are positioned in the southwestern corner of M33 and commented as `SW' in Table~\ref{tab:new_sg_m33}. To investigate if there are any supporting substructures for these potential supergiant members, we used an image from the {\it Galex} NUV band, as shown in the right panel of Fig.~\ref{fig:m33_structure}. An extended substructure originating from M33's disk is visible, marked by a white dashed box. Three YSG candidates are located on the edge of this substructure. As part of our future work, we plan to conduct follow-up observations for these candidates to further analyze this possible substructure.

\section{Summary}\label{sect:summary}
Based on LAMOST DR10 data, we conducted a systematic identification of supergiant members in M31 and M33. Firstly, objects of M31/M33 in LAMOST DR10 were cross-matched with the LGGS catalogue or {\it Gaia} DR3 to acquire their photometric data. BSG, YSG, and RSG candidates were then selected based on criteria adopted from previous studies. Subsequently, we excluded foreground field stars by comparing observed RVs to expected ones, as well as using the {\it Gaia} astrometry. In total, we identified 199 supergiant candidates in M31, with 168 `Rank1' and 31 `Rank2'. For M33, 84 supergiant candidates, including 56 `Rank1' and 28 `Rank2', were found. Our future work involves conducting follow-up spectroscopy for YSG candidates with low SNR to determine their memberships.

We constructed color-effective temperature and BC relations based on the YBC database to derive the effective temperatures and BCs of the candidates, and examined the distribution of these candidates in the HRDs. Furthermore, we checked the consistency between the locations of supergiant candidates and those expected from the Geneva evolutionary tracks. The results revealed an agreement, especially for `Rank1' candidates.

Upon analyzing the spatial distribution of candidates, we identified a potential substructure in the northeastern corner of M31 and another in the southwestern corner of M33. The northeastern corner of M31 suffers significant contamination from Galactic emission, posing challenges in determining whether the substructure belonging to M31 or primarily dominated by foreground emission. The follow-up validation of supergiant candidates within this substructure is crucial. Once confirmed, it would substantially contribute to ascertaining the ownership of the substructure. Likewise, the substructure in the southwestern corner of M33, along with the supergiant candidates within, also requires further analysis. These will be crucial aspects of our future research.

\begin{acknowledgements}
This work was funded by the National Natural Science Foundation of China (NSFC grant Nos. 11080922, 11803029, U1531244, 12090040, 12090044, 11833006, 12133001 and U1731308), Beijing Natural Science Foundation (no. 1242016) and the science research grants from the China Manned Space Project. This work was partially supported by Talents Program (24CE-YS-08) and the Popular Science Project (24CD012) of Beijing Academy of Science and Technology.

This work has made use of data products from the Guo Shou Jing Telescope (the Large Sky Area Multi-Object Fibre Spectroscopic Telescope, LAMOST). LAMOST is a National Major Scientific Project built by the Chinese Academy of Sciences. Funding for the project has been provided by the National Development and Reform Commission. LAMOST is operated and managed by the National Astronomical Observatories, Chinese Academy of Sciences.

\end{acknowledgements}

\newpage
\appendix                
\section{M31 and M33 supergiant candidates}
\setlength{\tabcolsep}{0.01mm}{
    \begin{ThreePartTable}
    \begin{longtable}[t]{cccccccccccc}
    \caption{ 199 supergiants in M31. A more detailed version, including RA, DEC, and LAMOST observation times with SNR $>$ 5, is available in electronic form in the online version of this manuscript (see beginning of Section~\ref{sect:result}).}\label{tab:new_sg_m31}  
    \\
    \hline
    Star$\,^{1}$ & $V_{\rm obs}$ &  $V_{\rm exp}$ & $V_{\rm obs} - V_{\rm exp}$ & $V$ & $B - V$ & SNR & Rank & Note$\,^{2}$ &Comment$\,^{3}$     \\ 
    \hline
    \endfirsthead
    \captionsetup{labelformat=empty,labelsep=none}
    \caption{Table A.1: continued}
    \\
    \hline
    Star$\,^{1}$ & $V_{\rm obs}$  &  $V_{\rm exp}$ & $V_{\rm obs} - V_{\rm exp}$ & $V$ & $B - V$ & SNR & Rank & Note$\,^{2}$ &Comment$\,^{3}$     \\ 
    \hline
    \endhead
    \hline
    \endfoot
    \hline
    \endlastfoot
    \multicolumn{10}{c}{Blue Supergiants} \\
    J0037+4016  & $-$450.1 & $-$497.7 & 47.6 & 15.854 & 0.231 & 25.7 & 1   & -- & Huang+2019; Liu+2022 \\
    J0037+4020  & $-$310.8 &$-$490.6 &179.8 &17.310 &0.107   & 10.1 & 1 & B1I & -- \\
    J0037+4021  & $-$546.7 & $-$498.3 & $-$48.3 & 18.146 & $-$0.002    & 7.9  & 1   & -- & --  \\   
    J0038+4032 &$-$462.8 &$-$477.6 &14.9 &17.763 &0.073  & 6.6 & 1  &B9.5I+ & -- \\
    J0039+4036 &$-$425.1 &$-$501.2 &76.1 &17.653 & 0.096&10.5 & 1   &B2.5Ia & -- \\
    J0039+4034  & $-$561.3 & $-$512.6 & $-$48.7 & 18.805 & 0.050    & 5.1 & 1    & -- &  \\
    J0039+4037  & $-$490.2 & $-$498.2 & 8.0 & 18.199 &  0.167   & 5.3 & 1    & -- & --  \\
    J0039+4045 &$-$431.9 &$-$461.2 & 29.3 &18.701 &$-$0.073 & 6.8 & 1  &O9.5I & -- \\
    J0039+4048 &$-$432.3 &$-$488.2 &15.8 &17.887 &0.126  & 7.6& 1  &B8I &-- \\
    J0039+4051 &$-$407.4 &$-$438.5 &31.1 &18.469 &$-$0.008 &8.1 & 1   &B8I & -- \\
    J0039+4051 &$-$565.8 &$-$439.6 &$-$126.1 &18.956 &$-$0.086 &5.5 & 1   &B1.5I & --\\
    J0039+4040  & $-$656.5 & $-$522.7 & $-$133.7 & 18.645 &  0.014   & 5.4 & 1    & --& -- \\
    J0039+4024 &$-$465.1 &$-$512.8 &47.7 &18.728 &0.014 &5.5 & 1  &B8Ia & --\\
    J0039+4040 &$-$470.8 &$-$527.7 &56.9 &18.250 &$-$0.090  & 11.6& 1  &B0.5I &--\\
    J0039+4023 &$-$598.9 &$-$507.4 &$-$91.5 &18.968 &0.095 & 5.3& 1  &B0.7Ia & --\\
    J0039+4028 &$-$513.0 &$-$522.2 &9.2 &18.338 &$-$0.014 & 5.5& 1  &B2.5I & --\\
    J0040+4027 & $-$568.2 & $-$518.1 & $-$50.1 & 18.718 &  $-$0.120   & 6.1 & 1    & --& -- \\
    J0040+4035 & $-$585.5 & $-$535.8 & $-$49.6 & 18.811 &  0.058   & 5.0  & 1   & --& -- \\
    J0040+4059 &$-$478.8 &$-$404.6 &$-$74.1 &18.595 &0.127 & 5.8& 1 &B2.5I &-- \\
    J0040+4032 & $-$515.3 & $-$523.0 & 14.7 & 17.901 & 0.055 & 12.5    & 1  & -- & Liu+2022\\ 
    J0040+4059 & $-$431.3 & $-$407.2 & $-$24.1 & 18.643 &   0.047  & 5.1 & 1  &-- &--  \\
    J0040+4045&$-$511.7 &$-$512.9 &1.2 &18.880 & 0.003& 5.4& 1 &B5I &-- \\
    J0040+4036&$-$570.0 &$-$536.0 &$-$34.0 &18.756 & $-$0.007&7.0 & 1 & B7I& --\\
    J0040+4031&$-$482.2 &$-$522.5 &40.4 & 18.590 &0.089 &6.1& 1 &B1I+Neb &-- \\
    J0040+4100  & $-$482.1 & $-$405.9 & $-$76.2 & 18.583 &  0.337   & 5.1 & 1   &-- &--   \\
    J0040+4029  & $-$472.9 & $-$512.4 & 39.5 & 18.846 &  $-$0.026   & 6.6  & 1  & --&--   \\
    J0040+4056 &$-$459.5 &$-$450.1 &$-$9.4 &17.967 &0.349  & 6.5& 1 &B8I &--\\
    J0040+4030 &$-$471.7 &$-$490.3 &18.6 &17.747 &0.111  & 11.8& 1 &BI &--\\
    J0040+4101  & $-$431.9 & $-$420.0 & $-$11.9 & 17.933 &  0.198   & 9.7 & 1   & -- &  Liu+2022 \\
    J0040+4030  & $-$365.1 & $-$469.9 & 104.8 & 18.412 &  0.049   & 9.5  & 1  & -- & --  \\
    J0040+4033&$-$439.5 &$-$473.8 &34.3 &16.989 & 0.216 &24.3 & 1 &cLBV &--\\
    J0040+4055 &$-$493.9 &$-$496.3 &2.4 &18.073 & 0.185 &8.9 & 1  &B2.5 &--\\
    J0041+4052 &$-$539.7 &$-$512.1 &$-$27.6 &17.485 &0.308 & 13.9 & 1  &A6I &-- \\
    J0041+4042  & $-$456.0 & $-$441.2 & $-$14.7 & 17.841 &  0.021   & 10.2  & 1  & -- &  Liu+2022 \\
    J0041+4119  & $-$280.8 & $-$317.9 & 37.1 & 18.617 &   0.606  & 7.3  & 1  & -- & -- \\
    J0041+4123 & $-$390.8 & $-$297.8 & $-$93.0 & 18.403 &  0.017   & 5.0 & 1  & --& -- \\
    J0041+4125 & $-$281.7 & $-$290.9 & 9.2 & 18.690 &  0.048   & 7.9  & 1  & --  &  -- \\
    J0041+4057 &$-$474.8 &$-$471.8 &$-$2.9 & 17.910 &0.168 & 11.3 & 1 &B8I &--\\
    J0042+4114  & $-$354.6 & $-$355.9 & 1.3 & 18.088 & 0.256 & 16.3  & 1   &  -- & Liu+2022 \\
    J0042+4136 &$-$253.7 &$-$257.9 &4.2 &17.807 & 0.056 &8.2 & 1  &B6V &--\\
    J0042+4135&$-$246.8 &$-$257.7 &10.9 &17.541 &0.225 &11.1 & 1  &B8I &-- \\
    J0042+4134 & $-$234.9 & $-$252.3 & 17.4 & 18.305 &  $-$0.088   & 5.3 & 1  & --& --\\
    J0042+4137  & $-$235.2 & $-$248.7 & 13.5 & 18.506 & 0.019 & 5.0   & 1   & -- & -- \\ 
    J0042+4056&$-$462.8 &$-$379.4 &$-$83.3 &18.430 & 0.026& 12.7 & 1 &B2I &-- \\
    J0042+4127 &$-$267.9 &$-$204.4 &$-$63.4 &17.265 & 0.012 & 10.4 & 1  &Star+N &--\\
    J0042+4048&$-$962.4 &$-$349.9 &$-$612.5 &18.142 &$-$0.048 &6.9 & 1  &B1.5I & --\\
    J0043+4103&$-$403.8 &$-$336.3 &$-$67.4 &18.386 & 0.278& 6.2 & 1  & B2I& -- \\
    J0043+4128  & $-$134.4 & $-$62.8 & $-$71.6 & 18.616 &  $-$0.049   & 10.3 & 1   & -- & -- \\
    J0043+4049 & $-$280.9 & $-$327.0 & 46.1 & 18.012 &  $-$0.008   & 6.4 & 1   & -- &  -- \\
    J0044+4117 & $-$272.8 & $-$244.4 & $-$28.4 & 18.788 &   0.367  & 11.6 & 1   & -- & --  \\
    J0044+4119 & $-$242.2 & $-$236.6 & $-$5.7 & 18.365 &  $-$0.067   & 9.6 & 1  & -- &  -- \\
    J0044+4204&$-$182.6 &$-$153.8 &$-$28.7 &16.465 &0.017 &14.1 & 1  & B0.5I& -- \\
    J0044+4155 &$-$275.5 &$-$112.2 &$-$163.3 &17.680 &$-$0.088  &7.0 & 1  &B1.5Ia &-- \\
    J0044+4153&$-$149.9 &$-$91.5 &$-$58.4 & 17.253&0.377 &9.8 & 1  &B8Ia & -- \\
    J0044+4200  & $-$222.7 & $-$111.0 & $-$111.7 & 17.878 &   $-$0.113  & 6.7  & 1  & --  & -- \\
    J0044+4129&$-$156.7 &$-$178.5 &21.8 &17.982 &$-$0.027 & 12.3 & 1  &B0.7Ib & -- \\
    J0045+4136&$-$132.8 &$-$135.2 &2.4 &17.962 &0.177 & 9.8& 1 & B8V&-- \\
    J0045+4156&$-$162.9 &$-$55.2&$-$107.6 & 17.108&0.068 &15.4 & 1  &B2.5I & --\\
    J0046+4215  & $-$147.2 & $-$71.3 & $-$75.8 & 18.218 &   $-$0.083  & 6.0 & 1  & -- &--   \\
    J0046+4138&$-$239.9 &$-$180.9 &$-$59.0 & 16.140 & 0.137&22.1 & 1  &B8Ie &-- \\
    LAMOST J0046+4232 &$-$188.6 & $-$96.7&$-$91.9 & -- & --  & 17.6 & 1 &-- & NE\\  
    J0047+4212 & $-$201.7 & $-$81.0 & $-$120.7 & 18.411 &$-$0.021     & 5.2  & 1  & --&  --\\
    \hline
    \multicolumn{10}{c}{Yellow Supergiants} \\
LAMOST J0036+3940 &$-$184.2&$-$534.3 &350.1 & --& --& 7.9 & 2 & -- &-- \\
LAMOST J0036+3931 &$-$471.9&$-$518.4 &46.5 & --& --& 14.5 & 1 & -- & --\\
LAMOST J0036+3956 &$-$510.2 &$-$534.2 &24.0 &-- &-- & 8.9 & 1 &-- &-- \\
J0037+4007 &$-$474.1 &$-$526.6 &52.5 & 18.474&0.496  & 5.7& 1 &YSG: &--\\
J0037+3954 & $-$274.3 &$-$535.8 &261.4 &16.789 &0.695 & 15.2 & 2 & fgd & --\\
J0037+3957 & $-$171.1 &$-$528.0 &356.9 &17.755 &0.438 & 5.6 & 1 & fgd & --\\
LAMOST J0037+3946  &$-$127.6&$-$521.9 &394.3 &-- &-- & 6.7 & 1 &-- &-- \\
J0037+3958&$-$509.0 &$-$535.0 &26.0 &17.160 &0.594  & 6.3& 1 & F5Ia& -- \\
J0037+4014 &$-$202.2 &$-$524.2 & 322.0 &16.135 & 0.935& 33.2 & 2 &fgd &-- \\
LAMOST J0037+4027 &$-$464.3 & $-$479.5&15.2 &-- &-- & 11.5 & 1 & -- & Weak 7774?\\
J0038+4032 & $-$179.3 & $-$473.3 & 294.0 &16.255 &0.666 & 19.4 & 2&fgd & --\\
J0038+4032 & $-$445.3 & $-$478.1 & 32.8 & 17.459 & 1.453 & 7.6  & 1 &-- & --\\
J0038+4008 & $-$228.8 & $-$533.8 & 305.0 & 16.324 & 0.692  & 6.9  & 1  & fgd & -- \\
J0038+4038 & $-$128.3 & $-$462.1 & 333.8 & 15.696 & 0.862  & 15.5  & 1  & -- & -- \\
J0038+3952 & $-$431.3 & $-$489.1 & 57.7 & 18.689 & 0.055 & 9.1  & 1 & -- & -- \\
J0039+4026 &$-$540.5 &$-$534.1 &$-$6.4 &16.735 &0.862  & 9.7 & 1 &G0Ia &-- \\
J0039+4020 &$-$511.8 &$-$529.4 &17.7 & 18.357&0.176 &6.1 & 1 &A0I & -- \\
J0039+4047 &$-$276.2 &$-$460.6 &184.3 &17.213 &0.603 &14.6 &1 & fgd &--\\
J0039+4024 & $-$398.6 & $-$529.8 & 131.2 & 18.562 &  0.511  & 5.6  & 1  & -- & --  \\
LAMOST J0039+3952 &$-$169.4 & $-$453.5&284.0 & -- & --  & 10.8 & 1 & -- & --  \\
J0039+4007&$-$362.8 &$-$482.8 &120.0 & 18.191& 0.801&5.1 & 2 &YSG: &--  \\
J0039+4040&$-$308.6 &$-$512.3 &203.7 &18.151 &0.542 &15.2 &1 & fgd &--\\
J0039+4051 & $-$440.2 & $-$439.3 & $-$0.9 & 17.267 & 1.471   & 7.7  & 1  &-- &-- \\
J0039+4035 &$-$502.1 &$-$532.9 &30.8 &17.170 &0.235 & 19.1 & 1  &YSG & 7774 \\
J0039+4043&$-$127.0 &$-$499.6 &327.4 &17.842 &0.829 &5.8 &1 & fgd &--\\
J0039+4038&$-$528.9 &$-$525.4 &$-$3.4 & 17.884&0.915 &7.0 & 1  &G8I & --\\
J0039+4038&$-$523.6 &$-$526.7 &3.0 &17.401 &0.653 &9.3 & 1 &F8I & --\\
LAMOST J0039+3946 &$-$131.2 &$-$436.1&304.9& --& --& 11.4 & 2 &-- & --\\
J0039+4053&$-$450.7 &$-$436.8 &$-$13.9 &17.344 &1.020 &13.4 & 1  &G4Ie & --\\
J0039+4028&$-$538.5 &$-$527.3 &$-$11.1 &17.451 &1.184 &5.7 & 1 & YSG&-- \\
J0039+4016 & $-$280.1 & $-$488.8 & 208.7 & 18.683 & 0.476   & 6.0  & 1  &-- & -- \\
J0040+4006&$-$294.8 &$-$457.4 &162.6 & 17.976&0.398 &7.4& 1  &YSG: & --\\
J0040+4103&$-$194.1 &$-$386.2 &192.1 &18.083 & 0.536& 12.3 & 1 &fgd &-- \\
J0040+4031&$-$572.8 &$-$523.8 &$-$49.0 &17.628 & 0.840&7.8& 1  &YSG & --\\
J0040+4037&$-$567.1 &$-$536.0 &$-$31.1 &18.272 & 0.203&8.3& 1  &A8I &-- \\
J0040+4044&$-$212.7 &$-$519.8 &307.2 &17.012 & 0.608&13.3& 1  &fgd &-- \\
J0040+4004&$-$449.6 &$-$440.7 &$-$8.9 & 18.334&0.256 &5.4 & 1&YSG  & --\\
LAMOST J0040+4112 &$-$148.9 &$-$348.8 &200.0 &-- & --&6.6 & 1 & --&-- \\
J0040+4031 &$-$532.0 &$-$509.0 &$-$23.0 &16.648 &0.275  & 16.2& 1 &F0Ia & 7774 \\
J0040+4056&$-$337.1&$-$441.9 &104.8&16.823&0.316 & 5.6& 2  &-- & --\\
J0040+4103&$-$163.3 &$-$397.9 &234.6 &18.089 & 0.361& 10.7 & 1 &fgd &-- \\
J0040+4027&$-$128.0 &$-$481.5 &353.5 &16.860 & 0.509& 6.1 & 1 &fgd &-- \\
J0040+4035 &$-$540.6 &$-$507.5 &$-$33.0 &17.701 &0.440 & 10.8& 1&F8I  &--\\
J0040+4058 & $-$187.0 & $-$438.4 & 251.3 & 18.517 &  0.537 & 7.8 & 1   &-- & --\\
J0040+4018 &$-$308.5 &$-$451.5 &143.0 &16.299 &0.614  & 17.1& 2 &YSG: &--\\
J0040+4007&$-$149.4 &$-$431.2 &281.9 &16.467 & 0.988& 12.8 & 1 &fgd &-- \\
J0040+4006 &$-$226.4 &$-$428.2 &201.8& 17.768 & 0.614 & 5.3 & 2 & -- &--\\
J0040+4017 & $-$208.4 & $-$443.0 & 234.6 & 17.306 &  0.436   & 15.1 & 1   & fgd &  -- \\
J0040+4057 & $-$175.9 & $-$462.0 & 286.1 & 17.344 &  0.573   & 9.1 & 1   & fgd &  -- \\
J0040+4045 & $-$300.9 & $-$533.4 & 232.4 & 16.554 & 0.258  & 8.6  & 1  & --& -- \\
J0040+4048 &$-$168.5 &$-$536.3 &367.8 &18.884 & 0.619& 9.5& 1& -- & -- \\
J0041+4013 &$-$133.5 &$-$412.5 & 279.0 &17.520 & 1.070 & 7.0 & 2 &fgd &-- \\
J0041+4104 & $-$192.0 & $-$434.4 & 242.4 & 18.543 & 0.587   & 13.5 & 1   &-- & --\\
LAMOST J0041+4124 &$-$157.9 &$-$306.0 &148.1 & --& --&21.6 & 2 &-- &-- \\
J0041+4018 &$-$136.7 &$-$413.7 &277.0 & 17.566& 0.786 &7.8 & 1 &fgd &-- \\
LAMOST J0041+4130 &$-$252.1 &$-$285.4 &33.3 &-- &-- &12.4& 2  & --& --\\
Mag J0041+4059 & $-$417.2 & $-$535.7 & 118.6 & 15.630 & 0.620  & 5.4  & 1  & --& --\\
J0041+4028&$-$337.2 &$-$401.1 &63.9 &18.291 &1.033 &15.7 & 2 &YSG &--\\
J0041+4111 & $-$296.6 & $-$395.8 & 99.1 & 18.465 & 0.615   & 9.3 & 1   &-- & -- \\
J0041+4127&$-$303.2 &$-$286.0 &$-$17.2 &17.994 &0.829 &9.3 & 1 &YSG &-- \\
J0041+4057 & $-$385.7 & $-$489.2 & 103.5 & 18.163 & 1.409  & 6.7 & 1   &-- & --\\
J0042+4102 & $-$498.1 & $-$493.1 & $-$5.0 & 18.829 &  0.112 & 9.2 & 1   & --& --\\
J0042+4051 &$-$444.4 &$-$412.9 &$-$31.6 &17.025 & 0.966 & 8.0& 1 &G5Ia &--\\
J0042+4051 &$-$452.5 &$-$411.9 &$-$40.6 &16.988 &0.761  & 5.6& 1 &F5Ia &--\\
J0042+4118 &$-$202.2 &$-$312.3 &110.1 &17.133 & 0.682& 12.8 & 1 &fgd &-- \\
J0042+4055 & $-$339.4 & $-$387.7 & 48.3 & 18.885 & 0.571   & 6.1 & 1  & --& -- \\
J0042+4106 &$-$420.7 &$-$425.0 &4.3 & 16.413& 0.750 &5.0 & 1 &fgd &-- \\
J0042+4057 &$-$448.9 &$-$384.7 &$-$64.2 &17.628 &0.425 &13.4& 1  &A4I & --\\
J0042+4121 &$-$235.1 &$-$265.7 &$-$30.6 &17.431 & 0.864& 9.7 & 1 &fgd &-- \\
J0042+4110 & $-$400.8 & $-$397.9 & $-$2.9 & 18.929 & 0.543   & 19.5 & 1   & --& -- \\
J0042+4129 & $-$276.0 & $-$235.3 & $-$40.7 & 18.209 & 0.259   & 10.1 & 1   &-- & --\\
J0042+4146 &$-$192.5 &$-$228.8 &36.3 &17.201 &0.739  & 6.6& 2 &YSG: & --\\
J0042+4032 &$-$235.2 &$-$361.4 &126.2 &17.788 & 0.459& 8.8 & 1 &fgd &-- \\
J0042+4144 &$-$223.8 &$-$226.8 &3.0 &17.985&0.181&5.5 & 1 &A7I & --\\
J0042+4144&$-$262.7 &$-$225.7 &$-$37.0 &16.410 & 0.465& 5.6 & 1 &fgd &-- \\
J0042+4138 &$-$141.7 &$-$224.1 & 82.4 &17.720 & 0.602& 8.9 & 1 &fgd &-- \\
J0042+4137 &$-$218.9 &$-$219.2 &0.3 & 14.975& 0.621 &35.8& 2  &YSG: &--\\
J0042+4145 & $-$222.1 & $-$220.9 & $-$1.2 & 17.632 &  0.161 & 5.2  & 1  & --& --\\
J0042+4130 & $-$181.3 & $-$210.0 & 28.7 & 18.753 & 0.606  & 5.6 & 1  &-- &-- \\
J0042+4132 &$-$255.0 &$-$213.0 & $-$42.0 &17.523 & 0.455& 7.7 & 1 &fgd &-- \\
LAMOST J0042+4153  &$-$149.5 &$-$349.5 &200.0 &-- & --&66.8 & 2 &-- & --\\
J0043+4103  & $-$375.1 & $-$336.7 & $-$38.4 & 18.719 &  0.654  & 10.7 & 1   & --&  --\\
J0043+4111 & $-$213.0 & $-$309.0 & 96.0 & 18.340 &  0.601 & 25.9  & 1  &-- & -- \\
J0043+4103&$-$377.1 &$-$332.0 &$-$45.1 &18.071 & 0.89&12.7& 1 &A5I & --\\
J0043+4109&$-$177.6 &$-$309.7 & 132.1 &17.304 & 1.332& 5.6 & 2 &fgd &-- \\
J0043+4141 &$-$184.0 &$-$194.5 &10.5 &16.973 &0.632  & 10.9& 1 &YSG: &--\\
J0043+4105 & $-$188.7 & $-$315.5 & 126.8 & 16.585 &  0.943 & 34.3 & 1   & fgd &  --\\
LAMOST J0043+4124 &$-$141.5 &$-$53.9 &$-$87.5 & --&-- & 7.9 & 1 &-- &-- \\
J0043+4153 & $-$171.6 & $-$201.6 & 30.0 & 16.541 &  0.781  & 8.1 & 1   & fgd &  --\\
J0043+4125 & $-$157.3 & $-$64.9 & $-$92.4 & 18.337 &  0.678  & 9.4 & 1   & --&  --\\
J0043+4122 & $-$180.2 & $-$132.4 & $-$47.8 & 18.373 & 0.549   & 16.2  & 1  &-- & -- \\
J0043+4119&$-$210.8 &$-$202.0 & $-$8.8  &15.467 & 1.012& 6.82 & 2 &fgd &-- \\
J0043+4110 & $-$271.6 & $-$287.8 & 16.2 & 18.458 & 0.340  & 14.5 & 1  & --& -- \\
J0043+4126 & $-$140.9 & $-$112.9 & $-$28.0 & 18.328 & 0.557 & 15.6  & 1  & --&--  \\
LAMOST J0043+4207 &$-$281.9 &$-$185.0 &$-$96.9 &-- & --& 31.2& 2  &-- &-- \\
LAMOST J0044+4056 &$-$131.5 &$-$310.1 &178.7 &-- & --& 5.3 & 1 &-- &-- \\
J0044+4158 & $-$236.4 & $-$163.5 & $-$72.9 & 18.856 & 0.480 & 7.6 & 1  & --& --\\
J0044+4201 &$-$206.2 &$-$165.3 &$-$40.9 &15.598 &0.462  &19.8& 1  &YSG: & Strong 7774 \\
J0044+4201 &$-$140.1 &$-$160.9 &20.9 &17.415 & 0.598& 8.3 & 2 & YSG:& --\\
Mag J0044+4116 & $-$228.9 & $-$252.9 & 24.0 & 15.150 &  1.180  & 21.6 & 1   & fgd &  --\\
J0044+4202 & $-$144.0 & $-$160.0 & 15.7 & 15.615 & 0.614   & 11.3 & 1   & --&  -- \\
J0044+4121 &$-$233.7 &$-$224.8 &$-$8.9 &16.727 &0.914  & 12.3& 1 &YSG: & Strong 7774 \\
J0044+4205 & $-$166.9 & $-$149.2 & $-$17.7 & 18.078 & 0.071 & 11.1 & 1   & --& -- \\
J0044+4135 & $-$146.2 & $-$84.7 &  $-$61.5 & 15.332 & 0.569 & 8.0 & 1   & --&-- \\
J0044+4133 & $-$132.5 & $-$116.6 & $-$15.9 & 17.429 & 0.639 & 5.8 & 1  & --&--  \\
J0044+4215 &$-$208.3 &$-$158.4 &$-$49.9 &17.355& 0.524 &10.1& 2  &-- & NE\\
J0044+4123 &$-$182.8 &$-$219.5 &36.7 &17.574 &0.652  & 7.3& 1 &YSG: &--\\
LAMOST J0044+4109 &$-$192.9 &$-$272.0 &79.1 & --& --&284.9 & 2& --& --\\
J0044+4214 & $-$152.9 & $-$142.8 & $-$10.1 & 18.705 & 0.043  & 5.5  & 1 & --& NE\\
J0045+4130 &$-$388.4 &$-$186.1 &$-$202.3 &16.834 &0.555  & 13.1& 1& YSG:&--\\
J0045+4226  & $-$193.5 & $-$144.0 & $-$50.0 & 17.096 &  0.787  & 9.1  & 1 & --&  NE \\
J0045+4132 &$-$488.8 &$-$187.7 &$-$261.1 &15.792 &0.853  &16.3& 1 &YSG: &--\\
J0045+4205 & $-$163.2 & $-$74.0 &  $-$89.2& 17.217 & 0.740  & 17.6  & 1 &-- &-- \\
J0045+4136 &$-$166.5 &$-$167.7 &1.3 & 15.877& 0.943&5.5& 1 &YSG: &--\\
J0045+4132 &$-$163.8 &$-$188.1 &24.3 &16.788 & 0.600&12.0 & 2& --& -- \\
J0045+4229 & $-$185.2 & $-$129.0 & $-$56.2 & 16.985 &  0.266 & 6.9 & 1  & --& NE \\
J0045+4158 & $-$236.4 & $-$54.8 & $-$181.6 &16.149 &0.647 & 30.3 & 2&-- & --\\
J0045+4133 & $-$377.8 & $-$190.5 & $-$187.2 &17.159 &0.592 & 6.2 & 2&fgd & --\\
J0046+4235 & $-$206.4 & $-$127.1 & $-$79.3 & 17.746 & 0.359  & 6.5  & 1 &-- & NE\\
J0046+4231 & $-$146.0 & $-$117.7 & $-$28.3 & 16.677 & 0.741  & 11.6 & 1  &-- & NE \\
J0046+4144 &$-$160.2 &$-$150.4 &$-$9.9 &15.224 & 0.654 & 27.6& 1& YSG:&--\\
LAMOST J0046+4214&$-$230.1 &$-$61.7 &$-$168.4 & --& --& 16.3& 2 & --&-- \\
LAMOST J0046+4209&$-$137.5 &$-$56.3 &$-$81.2 & --& --& 15.4 & 1& --& --\\
J0046+4208 & $-$150.8 & $-$58.8 & $-$92.0 &16.648 &0.467 & 7.5 & 2& --&-- \\
LAMOST J0047+4256 &$-$171.3 & $-$122.2&$-$49.0 &-- &-- &7.0 & 1 & --& NE\\
J0047+4212 & $-$168.5 & $-$66.6 &  $-$101.9 & 17.302 & 0.723  & 6.8  & 1 & --& --\\
J0047+4232 & $-$143.6 & $-$64.6 & $-$79.0 & 16.593 & 0.292  & 7.3 & 1 &-- &  NE\\
LAMOST J0048+4208&$-$136.2 &$-$106.6 &$-$29.6 &-- &-- & 15.1& 1 & --& --\\
LAMOST J0048+4238&$-$132.8 &$-$59.1 &$-$73.7 & --&-- & 25.0 & 1&-- & NE\\
J0048+4233 & $-$161.1 & $-$53.7 & $-$107.4 & 17.683 & 0.466  & 13.4 & 1  &-- & NE\\
LAMOST J0049+4210 &$-$104.9 &$-$143.4 &$-$38.5 & --& --& 49.0 & 2& --&-- \\
LAMOST J0049+4231 &$-$182.5 &$-$80.7 &$-$101.8 &-- &-- & 109.2 & 2& --&-- \\
LAMOST J0049+4252 &$-$155.0&$-$53.6 &$-$101.4 &-- & --& 9.05 & 2& --& -- \\
LAMOST J0049+4257 &$-$226.1&$-$53.7 &$-$172.4 & --& --& 13.9 & 2& --& -- \\
\hline
\multicolumn{10}{c}{Red Supergiants} \\
J0041+4110 &$-$390.0 &$-$407.1 &17.1 &19.626 &2.215 & 5.4& 1& RSG & Massey+2021 \\
J0043+4114 &$-$286.1 &$-$263.0 &$-$23.1 &19.935&2.438 & 5.7 & 1&RSG & Massey+2021\\
J0044+4155 &$-$136.3 &$-$116.9 &$-$19.3 &16.773 &2.046  & 5.1& 1 &fgd & -- \\
\hline
\end{longtable}
\begin{tablenotes}
    \item[1] Objects selected from LGGS are provided with LGGS names, while those from \cite{Magnier1992} or {\it Gaia} are labeled with Mag or LAMOST as a prefix before the coordinates.
    \item[2] 
     1) `Massey members' are provided with the spectral type from \cite{Massey2016}.\\
     2) `Massey non members' are noted as `fgd'.
    \item[3] 
    1) LBV identified by \cite{Huang2019} is added comment `Huang+2019'.\\
    2) BSGs analyzed by \cite{Liu2022} based on spectroscopic method are added comment `Liu+2022'.\\ 
    3) Candidates located within the substructure in the northeastern (southwestern) corner of M31 (M33) are added comment `NE (SW)'. \\
    4) RSGs identified by \cite{Massey2021} using NIR photometry are added comment `Massey+2021'.
    5) YSGs observed with O$\mathrm{I}$ $\lambda$7774 triplet in LAMOST spectra will be commented.
    \end{tablenotes}
\end{ThreePartTable}
}

\newpage
\setlength{\tabcolsep}{0.85mm}{
\begin{longtable}[t]{cccccccccc}
    \caption{84 supergiants in M33. A more detailed version, including RA, DEC, and LAMOST observation times with SNR $>$ 5, is available in electronic form in the online version of this manuscript.}\label{tab:new_sg_m33}  
    \\
    \hline
    Star &  $V_{\rm obs}$  &  $V_{\rm exp}$ & $V_{\rm obs} - V_{\rm exp}$ & $V$ & $B - V$ & SNR & Rank & Note &Comment     \\ 
    \hline
    \endfirsthead
    \captionsetup{labelformat=empty,labelsep=none}
    \caption{Table A.2: continued}
    \\
    \hline
    Star &  $V_{\rm obs}$  &  $V_{\rm exp}$ & $V_{\rm obs} - V_{\rm exp}$ & $V$ & $B - V$ & SNR & Rank & Note &Comment     \\ 
    \hline
    \endhead
    \hline
    \endfoot
    \multicolumn{10}{c}{Blue Supergiants} \\
J0132+3034  & $-$116.6  & $-$146.8  & 30.2 & 18.939  & $-$0.005 &  5.3  &1  & -- &-- \\ 
J0132+3024  & $-$140.5 & $-$115.8 & $-$24.7 & 15.406  & 0.258   &  17.0  &1   & --  &  Liu+2022\\
J0132+3038&$-$127.7 &$-$160.3 &32.6 &16.727 & 0.010 & 14.4 &1 &B3I & -- \\
J0132+3031&$-$121.7 &$-$127.3 & 5.6 &17.817  &0.135  & 6.1 &1  &-- &-- \\
J0133+3048&$-$118.4 &$-$202.5 &84.1 & 16.494& 0.113& 23.4 &1 &WNE+B3 &-- \\ 
J0133+3020&$-$151.1 &$-$105.1 &$-$46.0 &18.284 &$-$0.096 &6.0 &1 &B4I &-- \\
J0133+3030&$-$141.6 &$-$100.6 &$-$41.0 & 18.628 &$-$0.076 & 5.1 &1  & --&-- \\
J0133+3036&$-$152.4 &$-$100.1 &$-$52.3 & 16.904&0.031 &7.5 &1 &A6III &-- \\
J0133+3032&$-$153.2 &$-$116.7 &$-$36.4 & 17.194&0.034 &12.0 &1 &B9 &-- \\
J0133+3027 &$-$169.8 &$-$122.1 &$-$47.7 &17.613 &$-$0.049 &5.6 &1  & --&-- \\
J0133+3027&$-$184.9 &$-$125.5 &$-$59.4 &18.941 &0.023 &8.7&1 &F0I & --\\
J0133+3022 &$-$122.1 &$-$123.4 &1.3 & 18.657 &$-$0.051 &5.2 &1  & --&-- \\
J0133+3023&$-$146.7 &$-$123.5 &$-$23.2 &17.183 &0.099 &6.1 &1 &A0Ia+Neb & --\\
J0133+3023&$-$145.1 &$-$123.8 &$-$21.3 &16.395 & 0.009&12.6 &1 &B8Iae & --\\
J0134+3044  & $-$263.0 & $-$262.2  & $-$0.8 & 15.412  & 0.170 &   37.6  &1   & -- & Liu+2022 \\ 
J0134+3036 &$-$148.5 &$-$166.0 &17.5 & 18.145 &$-$0.018 & 9.7 &1 & -- & Liu+2022\\
J0134+3044 &$-$202.6 &$-$224.2 &41.6 & 18.166 & 0.086 &8.4 &1  & --&-- \\
J0134+3041&$-$166.9 &$-$217.1 &50.1 & 18.468&$-$0.160 & 7.2&1 & O9.5I& --\\
J0134+3033&$-$139.1 &$-$166.4 &27.3 &16.837 &0.131 &8.0&1 &cLBV &-- \\
J0134+3034&$-$150.1 &$-$172.7 &22.5 &18.969 &0.259 &7.7&1  &cLBV & --\\
J0134+3039&$-$171.3 &$-$201.6 &30.3 &17.143 &0.044 &21.3 &1  &B9I & --\\
J0134+3035&$-$156.2 &$-$179.4 &23.2 & 18.366&$-$0.059 &11.2&1 &B3I &-- \\
J0134+3049 &$-$327.8 &$-$252.3 &$-$75.5 & 18.381&$-$0.058 &11.7 &1  &-- &-- \\
J0134+3051  & $-$193.1 & $-$253.  & 60.6  & 18.622  & $-$0.05  &  6.7  &1   & -- & --\\ 
J0134+3046&$-$229.7 &$-$233.2 &3.5 & 17.296& $-$0.005&12.2&1  &B5I &-- \\
J0134+3027&$-$138.3 &$-$159.9 &21.6 & 18.069& 0.111& 16.3&1 &B8I&-- \\
J0134+3056&$-$246.1 &$-$259.8 &13.7 & 18.299&$-$0.021 & 7.2&1 &B2I &-- \\
J0135+3044&$-$143.2 &$-$214.8 &71.6 & 17.654& $-$0.038&13.6&1  & B5I& --\\

\hline
\multicolumn{10}{c}{Yellow Supergiants} \\
LAMOST J0131+3018 & $-$163.7 & $-$127.1 & $-$36.6 & -- & & 17.5 &2 & --&--\\
LAMOST J0131+3005 & $-$133.0 & $-$111.7 & $-$21.3 & -- &-- & 7.3 &1 &-- &SW\\
LAMOST J0131+3028 & $-$219.4 & $-$141.3 & $-$78.1 & -- &-- & 17.3&2  & --&--\\
LAMOST J0131+3001 & $-$189.5 & $-$105.5 & $-$84.0 &  --& --& 5.3 &2 &-- &SW\\
J0131+3032 & $-$179.6 &$-$148.2 & $-$31.4 & 18.557 & 0.553 & 6.8&2  & -- &--\\
LAMOST J0131+3041 & $-$126.0 & $-$166.7 & 40.7 & -- & --& 11.4 &2 & --&--\\
J0132+3006 &$-$149.1 & $-$105.7& $-$43.4&17.001  & 0.596 &11.6 &1 & -- & SW \\
J0132+3014 & $-$121.2 & $-$112.4 & $-$8.8&18.043 &0.629 &8.8 &2 & --& --\\
J0132+3033& $-$257.6&$-$148.9 &$-$108.7 &17.730  & 0.684 &7.6 &2  & fgd & --\\ 
J0132+3036& $-$221.3&$-$155.2 &$-$66.1 &17.532  & 0.580 &6.0 &2  & fgd & --\\ 
LAMOST J0132+3003 & $-$284.9 & $-$100.5 & $-$184.4 &  --& --& 7.4 &2 & --&SW\\
J0132+3022 & $-$137.5&$-$108.7 &$-$28.8 &18.831  & 0.069 &6.4 &1  & --& --\\
J0132+3045 &$-$130.5 &$-$182.3 &51.8 &14.483 &1.098 &49.7 &2  &YSG: &-- \\
J0132+3034 &$-$148.2 &$-$140.5 &$-$7.7 & 17.120& 0.141 &10.2 &1 &YSG & --\\
J0132+3029 &$-$131.3 &$-$123.0 &$-$8.4 & 17.143& 0.987&8.1 &2 &YSG:: &-- \\
J0133+3052 & $-$120.3&$-$204.9 &84.6 &17.436  & 0.543 &6.6 &2  & fgd & --\\ 
J0133+3037 &$-$161.8 &$-$151.1 &$-$10.6 &18.127 &1.230 &5.3 &1  &YSG & --\\
J0133+3034&$-$118.0 &$-$130.8 &12.8 &16.454&0.254 &10.3 &1 &F0Ia & --\\
J0133+3025&$-$127.6 &$-$100.4 &$-$27.2 & 17.032& 0.366& 27.3 &1 &YSG & Strong 7774\\
J0133+3104&$-$123.7 &$-$230.1 &106.4 & 17.694 &0.710 & 6.8 &2 & fgd & -- \\
J0133+3026&$-$138.1 &$-$100.4 &$-$37.6 & 15.583 &0.754 & 15.1 &2 & fgd & -- \\
J0133+3052 & $-$213.2 &$-$219.3 & 6.1 & 18.663 & 0.059 & 6.7 &1  & --&-- \\
J0133+3005 &$-$199.1 &$-$106.7 &$-$92.5 & 16.850& 0.610&5.1 &2  &YSG:: & --\\
J0133+3038 & $-$194.7& $-$155.1&$-$39.7 & 18.159 &0.726  &10.1 &1 & --& --\\
J0133+3057 &$-$202.1 &$-$231.1 &29.0 &18.703  &0.009 &5.6 &1  &-- &-- \\
J0133+3041 & $-$135.6 & $-$190.0 &54.4 &17.372 &0.899 & 11.4 &2  & --&--\\
J0133+3022&$-$164.9 &$-$106.0 &$-$58.9 &16.742&0.295 &5.6 &1 &YSG& --\\
J0133+3035&$-$130.4 &$-$105.3 &$-$25.2 &17.937 &0.062 &21.3 &1 &A5I & --\\
J0133+3031&$-$135.3 &$-$100.7 &$-$34.7 &16.201 &0.739 & 6.8 &1 &G0Ia & --\\
J0133+3052&$-$283.7 &$-$244.9 &$-$38.8 & 17.645&0.120 & 5.6 &1 &A4I & --\\
J0133+3100&$-$234.9 &$-$250.3 &15.4 &16.774 & 0.310& 9.3&1  &YSG & --\\
J0133+3009&$-$152.3 &$-$119.6 &$-$32.7 & 17.789& 0.551& 9.8&2 &YSG::& --\\
J0134+3050& $-$116.8 & $-$261.4 &144.6 & 17.029 &0.656 &26.3 &2 & --&--\\
J0134+3107&$-$166.4 &$-$255.0 &88.6 & 17.058&0.705 & 9.5 &2 &fgd &-- \\
J0134+3040&$-$156.5 &$-$208.8 &52.3 & 15.825 &1.073 & 6.8 &1 & fgd & -- \\
J0134+3046&$-$210.2 &$-$260.6 &50.4 & 15.996& 0.332& 13.1&1 &YSG & Strong 7774\\
J0134+3028&$-$173.2 &$-$144.4 &$-$28.8 & 17.284&0.854&20.4 &1 & F0Ie& --\\
J0134+3047 & $-$256.0 &$-$259.6 & 3.6 & 18.926 & 0.324 & 7.7&1  & --&--\\
J0134+3036& $-$168.2&$-$176.7& 8.4&18.722  &0.204  &9.0  &1 & --& --\\
J0134+3055&$-$220.3 &$-$263.7 &43.4 & 16.905& 0.246&10.5 &1 &YSG & Weak 7774 \\
J0134+3046&$-$263.5 &$-$252.4 &$-$11.0 & 16.868& 0.252& 7.4&1 &YSG & --\\
J0134+3057&$-$153.0 &$-$263.6 &110.6 & 17.037 &0.705 & 5.5 &2 & fgd & -- \\
J0134+3041&$-$165.4 &$-$211.5 &46.2 &16.765 &0.225 &5.6 &1 &A8Ia &-- \\
J0134+3038&$-$215.5 &$-$197.6 &$-$17.9 & 17.061& 0.286& 19.2&1 &YSG & 7774 \\
J0134+3026&$-$269.7 &$-$161.6 &$-$108.1 & 16.058&0.814 & 35.1&2 &fgd &-- \\
J0134+3054&$-$257.3 &$-$254.9 &$-$2.4 & 18.203&0.084 & 6.1&1 &YSG &-- \\
J0134+3041&$-$145.7 &$-$209.1 &63.4 & 16.468&0.854 &9.1 &2 &YSG:& --\\
J0135+3058&$-$135.3 &$-$254.1 &118.8 & 15.474&0.771 & 22.2&2 &fgd &-- \\
J0135+3105&$-$345.6 &$-$257.1 &$-$88.5 & 17.567&0.440 & 15.3&2 &fgd &-- \\
J0135+3105&$-$115.5 &$-$255.4 &139.9 & 15.963&0.567 & 8.6&2 &fgd &-- \\
J0135+3048&$-$112.9 &$-$223.2 &110.3 & 17.520&0.586 & 18.5&2 &fgd &-- \\
LAMOST J0135+3109 & $-$168.6 & $-$257.5 & 88.9 &  --&-- & 17.4 &2 &  --&--\\
LAMOST J0135+3058 & $-$147.3 & $-$239.7 & 92.4 & -- &-- & 25.4 &2& -- & --\\
\hline
\multicolumn{10}{c}{Red Supergiants} \\
J0133+3030&$-$127.8 &$-$114.0 &$-$13.7& 18.056& 2.061&6.5&1 &RSG&-- \\
J0133+3100&$-$198.9 &$-$222.8 &23.9 &16.002 & 1.621& 14.3&1&RSG &-- \\
J0134+3048&-263.5 &-262.0 &-1.5 &17.983 & 1.949&7.3 &1&M2-3Ia: & --\\
\hline
\end{longtable}    
}

\bibliographystyle{raa}
\bibliography{RAA-2024-0183}

\end{document}